\documentclass[journal]{IEEEtran}

\usepackage{amsfonts}
\usepackage{amssymb}
\usepackage{stfloats}
\usepackage{cite}
\usepackage{graphicx}
\usepackage{epstopdf}
\usepackage{psfrag}
\usepackage{subfigure}
\usepackage{amsmath}
\usepackage{array}
\usepackage{stfloats}
\usepackage{multirow}
\usepackage{color}
\usepackage{booktabs}
\usepackage{url}
\usepackage{graphicx}
\usepackage{enumitem}
\usepackage{lipsum} 

\usepackage{marvosym} 
\usepackage{url}
\usepackage{hyperref}
\usepackage{makecell}
\usepackage{bm}

\usepackage[linesnumbered,ruled,vlined]{algorithm2e}

\newtheorem{Thm}{Theorem}
\newtheorem{Lem}{Lemma}

\newtheorem{Def}{Definition}

\newtheorem{Prob}{Problem}
\newtheorem{Rem}{Remark}

\newtheorem{Cla}{Claim}

\begin{document}

\title{IREE Oriented Green 6G Networks: A Radial Basis Function Based Approach}

\author{Tao~Yu, Pengbo~Huang,  Shunqing~Zhang,~\IEEEmembership{Senior Member, IEEE,} Xiaojing~Chen,~\IEEEmembership{Member, IEEE,} Yanzan~Sun,~\IEEEmembership{Member, IEEE,} and Xin~Wang,~\IEEEmembership{Fellow, IEEE}
\thanks{ This work was supported by the National Key Research and Development Program of China under Grant 2022YFB2902304, the National Natural Science Foundation of China (NSFC) under Grants 62071284, the Innovation Program of Shanghai Municipal Science and Technology Commission under Grants 20JC1416400, 21ZR1422400, and 20511106603.}
\thanks{Tao Yu, Pengbo Huang, Shunqing Zhang, Xiaojing Chen, and Yanzan Sun are with Shanghai Institute for Advanced Communication and Data Science, Key laboratory of Specialty Fiber Optics and Optical Access Networks, Shanghai University, Shanghai, 200444, China (e-mails: \{yu\_tao, pbhuang, shunqing, jodiechen, yanzansun\}@shu.edu.cn).

Xin Wang is with the Key Laboratory for Information Science of Electromagnetic Waves (MoE), Department of Communication Science and Engineering, Fudan University, Shanghai 200433, China (e-mail: xwang11@fudan.edu.cn).}
\thanks{Corresponding Author: {\em Shunqing Zhang}.}
}
\markboth{Journal of \LaTeX\ Class Files,~Vol.~14, No.~8, August~2015}
{Shell \MakeLowercase{\textit{et al.}}: Bare Demo of IEEEtran.cls for IEEE Journals}
\maketitle

\begin{abstract}
In order to provide design guidelines for energy efficient 6G networks, we propose a novel radial basis function (RBF) based optimization framework to maximize the integrated relative energy efficiency (IREE) metric. Different from the conventional energy efficient optimization schemes, we maximize the transformed utility for any given IREE using spectrum efficiency oriented RBF network and gradually update the IREE metric using proposed Dinkelbach's algorithm. The existence and uniqueness properties of RBF networks are provided, and the convergence conditions of the entire framework are discussed as well. Through some numerical experiments, we show that the proposed IREE outperforms many existing SE or EE oriented designs and find a new Jensen-Shannon (JS) divergence constrained region, which behaves differently from the conventional EE-SE region. Meanwhile, by studying IREE-SE trade-offs under different traffic requirements, we suggest that network operators shall spend more efforts to balance the distributions of traffic demands and network capacities in order to improve the IREE performance, especially when the spatial variations of the traffic distribution are significant. 
\end{abstract}


\section{Introduction} \label{sect:intro}
\IEEEPARstart{B}{it}-per-Joule energy efficiency (EE) metric has been proposed, investigated, and standardized during the past two decades \cite{zhang2019first}, which links transmission capability and energy consumption together to reflect the energy utilization of many wireless communication systems. Guided by this bit-per-Joule EE metric, a variety of energy efficient solutions have been extensively studied. For example, heterogeneous network \cite{badic2009energy} has been recognized as an energy efficient network architecture to achieve differentiated traffic coverage, and massive multiple-input-multiple-output schemes \cite{lu2014overview} have been developed to improve the EE performance for more than 100 times \cite{bjornson2014massive}. 

With the goal to intelligently connect everything via nearly unlimited wireless resources, the concept of sixth-generation (6G) wireless communication networks \cite{dang2020should} has been widely discussed recently. For instance, Terahertz \cite{8663550} and visible light communication \cite{matheus2019visible} have been shown to be promising for Tbps throughput, along with ultra-scale multiple antennas and over tens of gigahertz (GHz) available bandwidth. Space-air-ground integrated network (SAGIN) \cite{liu2018space-air-ground} and intelligent reflecting surface (IRS) \cite{huang2019reconfigurable} have been proposed to reconstruct the wireless propagation environment with three dimensional coverage for more energy efficient and reliable connections. By applying the conventional EE metric, the above technologies can be further optimized towards the energy efficient direction. For example, by simultaneously optimizing transmit power allocation and phase shifts, IRS-based resource allocation methods can achieve up to $300\%$ higher EE than conventional multi-antenna amplify-and-forward relaying scheme \cite{huang2019reconfigurable}. For the mmWave and sub-THz indoor system, a higher EE is guaranteed with an improved antenna arrays structure design \cite{lin2016energy}.
However, all the above energy efficient designs rely on the full traffic load assumption, and the unbalanced traffic distribution in the practical network deployment has been rarely considered.

To address this issue, we have proposed a novel EE metric named {\em integrated relative energy efficiency} (IREE) in our previous work \cite{yu2022novel}, which incorporates the non-uniform traffic distribution in the EE evaluation via the famous Jensen-Shannon (JS) divergence \cite{manning1999foundations}. By measuring the statistical mismatch between the network capacity and the traffic distribution, the IREE metric is able to incorporate the network capability and the dynamic traffic variations together for a more effective EE evaluation. However, we cannot straight-forwardly extend the existing energy efficient optimization approaches to the IREE metric due to the following reasons. First, the existing energy efficient optimization frameworks rely on the deterministic traffic assumption, which is insufficient to deal with the non-uniform and time-varying traffic distributions. Second, to obtain the explicit expressions of IREE is also challenging, since the interactions between the instantaneous traffic and network capacity are difficult to describe. Last but not least, the systematical approach to approximate the IREE based performance metric with general traffic and capacity distributions is still open in the literature.

In order to provide the design guidelines for energy efficient 6G networks, we deal with the above issues by proposing a radial basis function (RBF) based IREE maximization framework in this paper. Specifically, we approximate the non-uniform and time-varying traffic  using a novel RBF neural network, where each neuron is designed based on the spectral efficiency (SE) oriented RBF. Together with the Dinkelbach's algorithm, the proposed IREE maximization scheme is able to iteratively find the optimized value of IREE and the corresponding resource allocation strategy. In addition, we discuss the potential issues in the training process and propose a two-stage training strategy accordingly. Based on that, we analyze the convergence properties of the entire IREE maximization scheme as well. Through some numerical results, we show that the proposed IREE maximization scheme is able to achieve 
$123.0\% \sim 185.9\%$ IREE improvement if compared with some conventional EE oriented designs, and guarantee different traffic requirement with different distributions.

\subsection{Related Works} \label{subsect:related}

To improve the conventional bit-per-Joule EE performance, a variety of network deployment strategies have been proposed over the past few decades. For example, a multi-objective genetic algorithm has been proposed in \cite{dai2020propagation} to reduce the number of base stations (BSs) and improve the network coverage performance for given network traffic profile, respectively. In order to fit the daily traffic variations, the BS sleeping technology with cell zooming has been proposed in \cite{sharma2019transfer}, which adapts the network topology in a large timescale. In a small timescale, however, different types of shutoff strategies have been widely used to save the power and improve the EE performance. For instance, the symbol level shutoff scheme has been proposed in \cite{chen2011network} to save the dynamic power consumption of power amplifiers (PAs). The similar idea has been extended to the small signal circuit level and the carrier level in \cite{frenger2019more} and \cite{piovesan2022machine}, respectively, which shows promising EE improvement under practical network environment. Recently, machine learning (ML) based EE schemes have been proposed as well, which can be generally classified into four categories, including deployment optimization \cite{dai2020propagation, moysen2016machine}, mode selection  \cite{donevski2019neural, liu2018deepnap}, user association \cite{wang2019reinforcement} and power allocation \cite{9286851, 9109742}. In the above EE schemes, deep neural network and convolutional neural network architectures are commonly employed for channel characteristic extraction, and the recurrent neural network and deep Q network are often employed to deal with time-varying traffics. All the above ML-based designs typically treat neural networks as ``black boxes'', lacking physical interpretability.

With the vision to intelligently connect almost everything in our world, the design principle of 6G has been switched to support sufficient traffic demand over the entire earth surface through advanced network architecture and transmission technologies \cite{dang2020should}. Since the guarantee of coverage and quality of experience (QoE) over the heterogeneous spaces is critical to the success of 6G \cite{hossfeld2023greener}, the deployment of conventional energy efficient schemes becomes challenging. For example, different levels of BS sleeping/shutoff schemes may not be preferable due to the moving coverage of SAGIN BSs and the limited circuit power consumption of IRS architectures \cite{zhu2022creating} \cite{liu2021intelligent}. In order to meet the service requirement for 6G networks while preserving the energy consumption, many energy efficient multi-band management schemes have been proposed \cite{saeidi2023multi} to deal with massive available bandwidth in millimeter wave or Terahertz bands. The EE properties of IRS based architectures have been discussed in \cite{liu2021intelligent} to provide guidelines for the energy efficient passive beam management. However, all the aforementioned bit-per-Joule EE oriented designs focus on improving the network capability, and the heterogeneity of wireless services in terms of non-uniform traffic distribution and unbalanced variations is rarely considered.

In order to address the above issues and provide more valuable design insight, we have proposed a novel IREE metric \cite{yu2022novel} to capture the mismatch between wireless capacity and traffic requirement distributions through the JS divergence. Neither the conventional EE oriented designs \cite{7446253} nor the variation inference method \cite{blei2017variational} can be applied to simultaneously maximize the EE performance and minimize the JS divergence, and a new approach to minimize the IREE is thus required.

\subsection{Contributions \& Organizations} \label{subsect:contribution}

In this paper, we propose an IREE oriented green 6G network design framework using a RBF based approach. The main contributions are summarized as follows.

\begin{itemize}
     \item{\em IREE Maximization Framework.} 
     In order to obtain the optimal IREE, we analyze the optimal condition for the original IREE maximization problem and derive an iterative solution thereafter. Specifically, we propose a RBF network optimization scheme to allocate resources for any given IREE, and gradually update the IREE value using proposed Dinkelbach's algorithm accordingly. We then obtain the convergence conditions for the proposed IREE maximization framework with some mathematical proofs and compare it with other conventional EE oriented design to show the benefits via numerical examples.
    \item{\em SE based RBF Network Design.} 
    Within the IREE maximization framework, we design an interpretable SE based RBF network by providing the existence and uniqueness conditions accordingly. Specifically, we model wireless networks using RBF based networks, where each RBF neuron represents the SE of each BS. We then derive the corresponding two-stage training strategies with provable convergence properties and the associated optimality gap. Through the proposed SE based RBF network design scheme, we can approximate any continuous traffic distributions, and describe the interactions between the instantaneous traffic and network capacity by data driven approaches. On top of that, we are able to characterize the relationship between total network power consumption, network capacity, and JS divergence, thereby achieving the optimized IREE.
    \item {\em IREE-SE Trade-offs and Design Principle.} Based on the proposed IREE maximization scheme, we numerically study the trade-off relations between IREE and SE, and characterise different operation regions by varying the network capacity distributions. Different from the conventional EE-SE trade-offs as illustrated in \cite{zhang2016fundamental}, the IREE-SE trade-off introduces {\em the JS divergence constrained region}, where the IREE varies and the SE remains unchanged. By studying different IREE-SE trade-off curves under different traffic requirements, we conclude that the network operators shall spend some efforts to balance the traffic demand and the network capacity distribution in order to improve the IREE performance, especially when the spatial variations of the traffic distribution are significant (e.g., in the urban scenario).
\end{itemize}

The remainder of this paper is organized as follows. We provide the system model and formulate the IREE maximization problem in Section~\ref{sect:system_model}. A novel IREE maximization scheme is proposed in Section~\ref{sect:proposed_scheme}, followed by some training skills and performance analysis in Section~\ref{sect:training}. In Section~\ref{sect:num_res}, we provide some numerical examples on the proposed IREE maximization scheme. Finally, concluding remarks are provided in Section~\ref{sect:conc}.

\section{System Models and Problem Formulation} \label{sect:system_model} 

In this section, we briefly introduce the wireless network and traffic models adopted in this paper, and formulate the IREE maximization problem in what follows.

\subsection{System Model}

\begin{figure}[t] 
\centering  
\includegraphics[height=7cm,width=8.5cm]{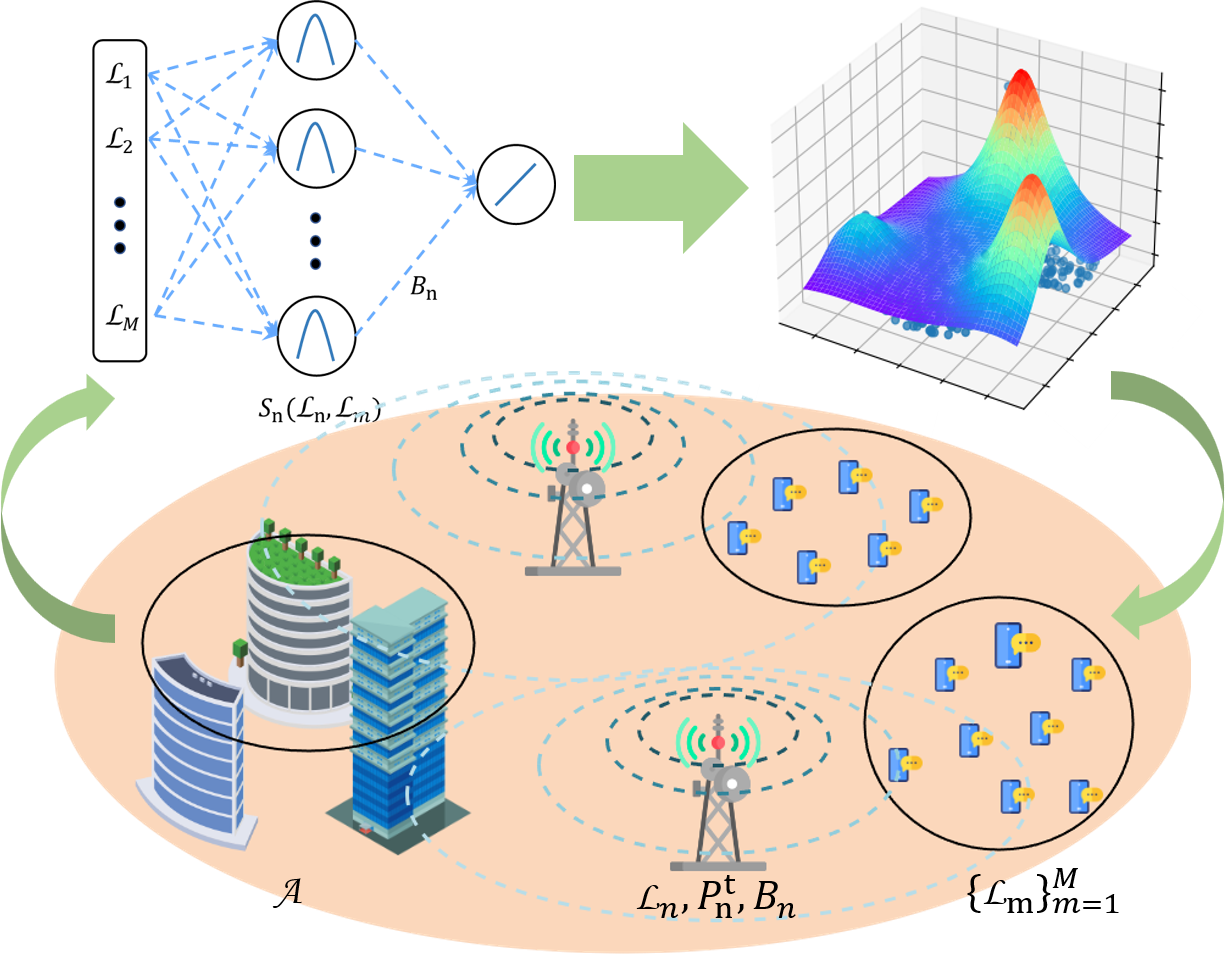}
\caption{An illustrative example of wireless networks and traffics within area $\mathcal{A}$. The $N_{BS}$ deployed BSs can be regarded as radial basis neurons, which are used to fit any continuous traffic distribution, thereby reducing JS divergence while increasing network capacity.}
\label{fig:scenario}
\end{figure}

Consider a wireless communication network with $N_{BS}$ BSs as shown in Fig.~\ref{fig:scenario}, where a total bandwidth of $B_{\max}$ are shared without overlap. Denote $\mathcal{L}_n$ and $B_n$ as the location and the available transmission bandwidth of the $n^{th}$ BS, and for any receiving entity with location $\mathcal{L}$, the normalized path loss coefficients with respect to the $n^{th}$ BS, $L(\mathcal{L}, \mathcal{L}_n)$, can be obtained via \cite{ku2013spectral},
\begin{equation} 
\label{eq:pathloss}
L(\mathcal{L}, \mathcal{L}_n) = \gamma ||\mathcal{L} - \mathcal{L}_n||_2^{\alpha} + \beta, \quad \forall n \in [1,\ldots, N_{BS}],
\end{equation}
where $\alpha > 2 $ denotes the path loss exponent, $\beta > 0$ and $\gamma > 0$ represent the normalization factors. According to the Shannon's capacity theorem \cite{shannon1948mathematical}, the equivalent capacity expression from the $n^{th}$ BS is thus given by\footnote{For illustration purpose, we consider the point-to-point unicast case in the following derivation. However, the proposed framework can be straight forwardly extended to OFDMA based systems \cite{6334508} by replacing $B_n$ with different sub-carrier spacings.},
\begin{equation} 
\label{eq:capacity}
C_n(\mathcal{L}) = B_n \log_2 \left (1 + \frac{ P^t_n/L(\mathcal{L}, \mathcal{L}_n)}{ \sigma^2 B_n} \right),
\end{equation}
where $P^t_n$ denotes the transmit power of the $n^{th}$ BS and $\sigma^2$ represents the power spectrum density of the additive white Gaussian noise. By summarizing over all $N_{BS}$ BSs, the total capacity at the location $\mathcal{L}$, $C_T(\mathcal{L})$, can be obtained by,
\begin{eqnarray}
C_T(\mathcal{L}) & = & \sum_{n=1}^{N_{BS}} C_n(\mathcal{L}). \label{eqn:tot_cap}
\end{eqnarray}

In order to transmit power $P^t_n$ in the air interface, the entire power consumption of the $n^{th}$ BS is given by $\lambda P^{t}_n + P^{c}$ \cite{6334508}, where $\lambda$ denotes the power amplify coefficient and $P^{c}$ denotes the static circuit power. Therefore, the total amount of power consumption $P_T$ is given by,
\begin{eqnarray}
\label{eqn:total_pow_def}
P_T & = & \sum_{n=1}^{N_{BS}} \left(\lambda P^{t}_n + P^{c}\right).
\end{eqnarray}

The following assumptions are adopted throughout the rest of this paper. First, the total amount of traffic requirement at the location $\mathcal{L}$ is given by $D_T(\mathcal{L})$. Second, $\beta, \gamma, \lambda$, and $P^{c}$ are assumed to be constant during the evaluation period\footnote{The non-constant channel fading effects, such as shadowing, will be discussed through numerical results in Section.~\ref{subsec:design_prin}.}. Last but not least, the target evaluation area is denoted by $\mathcal{A}$ and the distribution of $D_T(\mathcal{L})$ is assumed to be continuous over the entire area  $\mathcal{A}$.

\subsection{Problem Formulation}
\label{subsec:formu}
With the above mathematical models, we can have the IREE definition by incorporating the mismatch between network capacities and traffic requirements as follows.

\begin{Def}[IREE Metric \cite{yu2022novel}] \label{def:IREE}
The IREE of wireless networks, $\eta_{IREE}$, is defined to be,
\begin{eqnarray} \label{eq:def_iree}
\eta_{IREE} = \frac{\min\{C_{Tot},D_{Tot}\}\left[1 - \xi\left(C_{T}, D_{T}\right)\right]}{P_T}.
\end{eqnarray}
In the above expression, $C_{Tot} = \iint_{\mathcal{A}}C_{T}(\mathcal{L}) \textrm{d}\mathcal{L}$ and $D_{Tot} = \iint_{\mathcal{A}}D_{T}(\mathcal{L}) \textrm{d}\mathcal{L}$ denote the total amount of wireless capacity and the total amount of wireless traffic over the entire 2D area $\mathcal{A}$. $\xi\left(C_{T}, D_{T}\right)$ is the JS divergence \cite{manning1999foundations} as given in \eqref{eqn:js_divergence}.
\end{Def}

\begin{figure*}[t]
\begin{eqnarray}
 \label{eqn:js_divergence}
 \xi\left(C_{T}, D_{T}\right) = \frac{1}{2} \iint_{\mathcal{A}}  \frac{C_{T}(\mathcal{L})}{C_{Tot}} \log_2 \left[ \frac{2 D_{Tot} C_{T}(\mathcal{L}) }{ D_{Tot} C_{T}(\mathcal{L}) +  C_{Tot} D_{T}(\mathcal{L}) } \right] + \frac{D_{T}(\mathcal{L})}{D_{Tot}} \log_2 \left[ \frac{2  C_{Tot} D_{T}(\mathcal{L}) }{ C_{Tot} D_{T}(\mathcal{L}) + D_{Tot} C_{T}(\mathcal{L}) } \right] \textrm{d}\mathcal{L}.
 \end{eqnarray}
\end{figure*}

As a result, the IREE oriented schemes can consider both network capacity improvement and traffic mismatch simultaneously, and we formulate the IREE maximization problem as below.

\begin{Prob}[Original IREE Maximization Problem] \label{prob:origin}
The IREE of the wireless communication network with $N_{BS}$ BSs can be maximized by the following optimization problem.
\begin{eqnarray}
    \underset{\{\mathcal{L}_n\}, \{B_n\},\{P^t_n\}}{\textrm{maximize}} && \eta_{IREE}, \nonumber \\
    \textrm{subject to} && \eqref{eq:pathloss} - \eqref{eqn:js_divergence}, \nonumber \\
    && \zeta(\{\mathcal{L}_n\}, \{B_n\},\{P^t_n\}) \geq \zeta_{\min}, \label{constrain:qos} \\
    && \sum_{n=1}^{N_{BS}} B_n \leq B_{\max}, \forall B_n \geq 0, \\
    && \sum_{n=1}^{N_{BS}} P^t_n \leq P_{\max}, \forall P^t_n \geq 0.  \label{constrain:max_power} 
\end{eqnarray} 
In the above mathematical problem, $B_{\max}$ and $P_{\max}$ denote the total bandwidth and power limit, respectively. Note $\zeta(\{\mathcal{L}_n\}, \{B_n\},\{P^t_n\}) = \frac{\min\{C_{Tot},D_{Tot}\}\left[1 - \xi\left(C_{T}, D_{T}\right)\right]}{D_{Tot}}$ as the network utility indicator, the constraint \eqref{constrain:qos} ensures the minimum traffic requirement with $\zeta_{\min}  \in [0,1]$.
\end{Prob}

Problem~\ref{prob:origin} is in general difficult to solve due to the following reasons. First, the explicit expression of the IREE cannot be easily obtained, since a priori knowledge of $D_{T}(\mathcal{L})$ is still missing. Second, $\xi\left(C_{T}, D_{T}\right)$ is strictly non-convex with respect to the optimizing parameters  $\{\mathcal{L}_n\}, \{B_n\}$ and $\{P^t_n\}$. Last but not least, even if some iteration based algorithm, such as iterative water-filling method \cite{cheng1993gaussian}, could be applied, the uniqueness and convergence properties are yet to be analyzed.

\section{Proposed IREE Maximization Scheme} \label{sect:proposed_scheme} 

In this section, we rely on the Dinkelbach's algorithm \cite{dinkelbach1967nonlinear} to maximize the IREE metric, where the original fractional programming problem is transformed into the general non-convex optimization problem for any given $\eta_{IREE}$. Through the proposed RBF architecture, we are able to obtain the optimal strategy $\{\mathcal{L}_n^{\star}\}, \{B_n^{\star}\},\{P^{t,\star}_n\}$ accordingly.

Denote $\eta_{IREE}^{\star}$ to be the optimal value of Problem~\ref{prob:origin}, we can rewrite the original IREE maximization problem into linear forms, and according to \cite{schaible1976fractional}, we have the following lemma for $\eta_{IREE}^{\star}$.

\begin{Lem}[Optimal Condition] \label{lem:optimal_condition} The optimal value of IREE, i.e., $\eta^{\star}_{IREE}$, can be achieved if and only if the following equation is satisfied.
\begin{eqnarray} \label{eq:opt}
\underset{\{\mathcal{L}_n\}, \{B_n\},\{P^t_n\}}{\textrm{maximize}}  \Big\{ \min\{C_{Tot},D_{Tot}\}\big[1 - \nonumber \\
\xi\left(C_{T}, D_{T}\right)\big]  
- \eta^{\star}_{IREE} P_T \Big\} = 0,
\end{eqnarray}
where $\{\mathcal{L}_n\}, \{B_n\},\{P^t_n\}$ satisfy \eqref{eq:pathloss} - \eqref{constrain:max_power}.
\end{Lem}

By applying Lemma~\ref{lem:optimal_condition}, we can decouple the original IREE maximization problem into the following two procedures, i.e., we solve the IREE maximization problem for given $\eta_{IREE}$ in the first procedure, and adopt the Dinkelbach's algorithm \cite{dinkelbach1967nonlinear} to update $\eta_{IREE}$ in the second procedure. The above two procedures are running iteratively to find the optimized $\eta^{\star}_{IREE}$. With the above illustration, we can solve the following utility maximization problem for given IREE rather than the original IREE maximization problem to reduce the computational complexity.

\begin{Prob}[Utility Maximization for Given IREE] 
\label{prob:transformed}
For any given IREE, the utility function, $\min\{C_{Tot},D_{Tot}\}\big[1 - \xi(C_{T}(\mathcal{L}), D_{T}(\mathcal{L})) \big] - \eta_{IREE} P_T$, can be maximized via the following optimization problem.
\begin{eqnarray}
    \underset{\{\mathcal{L}_n\}, \{B_n\},\{P^t_n\}}{\textrm{maximize}} && \min\{C_{Tot},D_{Tot}\}\left[1 - \xi\left(C_{T}, D_{T}\right)\right] \nonumber \\
    && -\eta_{IREE} P_T, \\
    \textrm{subject to} && \eqref{eq:pathloss} - \eqref{constrain:max_power}. \nonumber 
\end{eqnarray}
\end{Prob}

Problem~\ref{prob:transformed} is also challenging, since the aforementioned three issues of Problem~\ref{prob:origin} are still valid. To make it mathematical tractable, we propose to design RBF based on the Shannon's capacity formula \cite{shannon1948mathematical}, and generate the corresponding neural networks to approximate the continuous capacity distribution of $C_{T}(\mathcal{L})$. Through this approach, we can numerically obtain the exact value of $\xi\left(C_{T}, D_{T}\right)$ for any given continuous traffic distribution of $D_{T}(\mathcal{L})$, which can be minimized using data driven approaches. In the following parts, we focus on introducing the spectrum efficiency (SE) based RBF and discussing the uniqueness and convergence properties.

\IncMargin{1em}
\begin{algorithm} [t] 
\label{alg:Dinkelbach}  
\caption{Proposed IREE Maximization Scheme} 
\SetKwInOut{Input}{input}
\SetKwInOut{Output}{output}
\SetKwRepeat{Do}{do}{while}
	
\Input{ $ D_T(\mathcal{L}) $, $N_{BS}$, $\alpha$, $\beta$, $\gamma$, $\{ \lambda \}_{n=1}^{N_{BS}}$, $\{ P^{c} \}_{n=1}^{N_{BS}}$, $B_{\max}$, $P_{\max}$ } 

\Output{$\{\mathcal{L}_n^{\star}\}, \{B_n^{\star}\},\{P^{t,\star}_n\}, \eta^{\star}_{IREE}$}

\BlankLine 

Initialization: $k=1$, $\epsilon >0$, $\eta_{IREE}^{(1)}$, $ \{ \mathcal{L}_n^{(0)}\}, \{B_n^{(0)}\},\{P^{t,(0)}_n\}$; 

\While{$ | L_{err}^{(k)}(\eta^{(k)}_{IREE}) | > \epsilon$ } {  

Solve Problem~\ref{prob:transformed} using the proposed RBF network and Adam optimizer, $\{\mathcal{L}_{n}^{(k)}\}, \{B_{n}^{(k)}\},\{P^{t,(k)}_{n}\} = \arg \underset{\{\mathcal{L}_n\}, \{B_n\},\{P^t_n\}}{\textrm{minimize}} L_{err}^{(k)}( \{\mathcal{L}_n\}, \{B_n\},\{P^t_n\} ;\bm{\omega} )$

Update IREE according to \eqref{eqn:iree_update} and obtain $\eta^{(k+1)}_{IREE}$;

$k = k+1$;

}

Optimized IREE and parameters:  $\eta^{\star}_{IREE} = \eta^{(k)}_{IREE} $, $\{\mathcal{L}_n^{\star}\}, \{B_n^{\star}\},\{P^{t,\star}_n\} = \{\mathcal{L}_{n}^{(k-1)}\}, \{B_{n}^{(k-1)}\},\{P^{t,(k-1)}_{n}\}$;
\end{algorithm}
\DecMargin{1em} 

\subsection{SE based RBF Design}

\begin{figure*}[t] 
\centering  
\includegraphics[height=7cm,width=17.5cm]{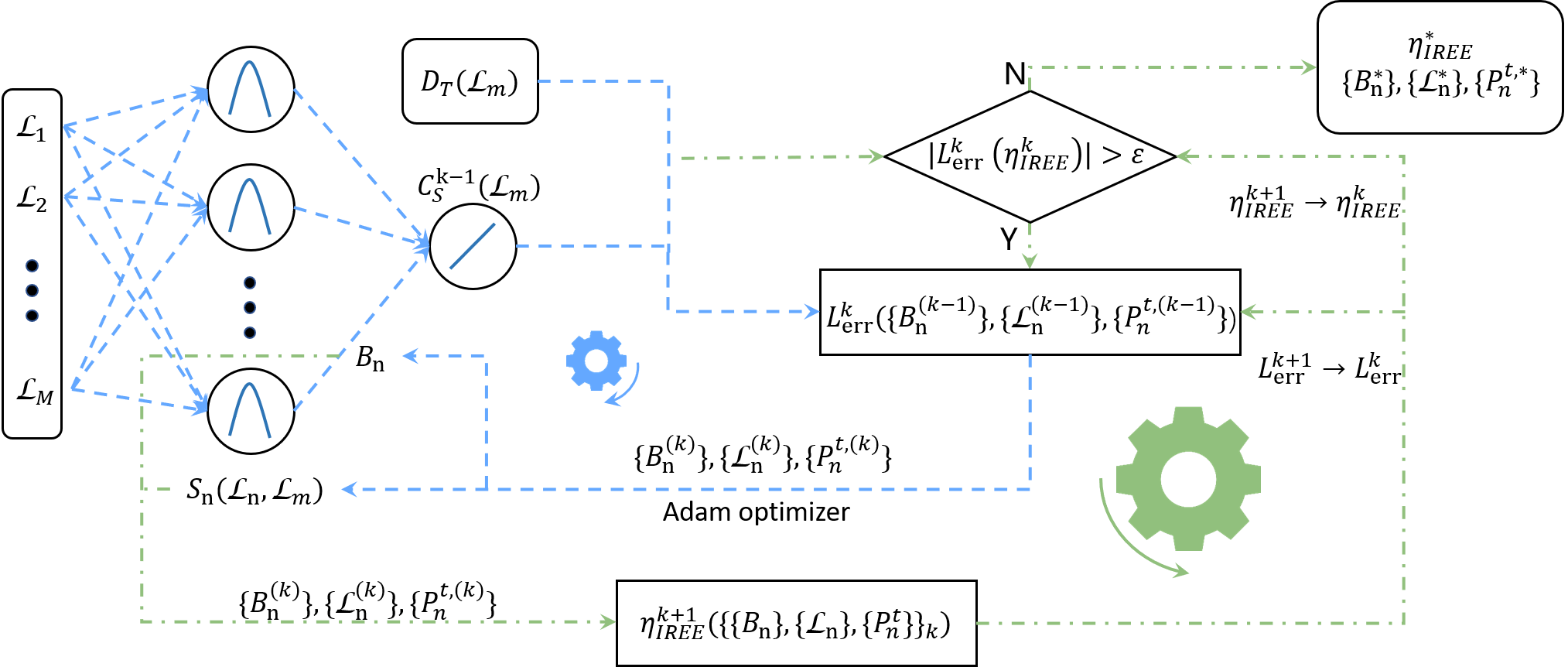}
\caption{An illustration of proposed scheme, where the blue gear is the driving gear and represents the forward and backward propagation through the RBF network to minimize $L_{err}^{(k)}$ and thus achieve better IREE. The green gear is the driven gear and represents to obtain the optimized IREE through a series of $L_{err}^{(k)}$ minimization problems, where $L_{err}^{(k)}$ is constructed through IREE in current iteration $\eta_{IREE}^{(k)}$. }
\label{fig:algorithm}
\end{figure*}

By combining equations \eqref{eq:capacity} and \eqref{eqn:tot_cap}, we can have a lower bound of the capacity $C_T(\mathcal{L})$ as given by,
\begin{eqnarray}
\label{eqn:def_Cs}
C_T(\mathcal{L}) &=& \sum_{n=1}^{N_{BS}} B_n \log_2 \left (1 + \frac{ P^t_n/L(\mathcal{L}, \mathcal{L}_n)}{ \sigma^2 B_n} \right), \nonumber \\
&\geq& \sum_{n=1}^{N_{BS}} B_n S_n(\mathcal{L}_n, \mathcal{L}) \triangleq C_S(\mathcal{L}).
\end{eqnarray}
In the above equation, $S_n(\mathcal{L}_n, \mathcal{L})$ denotes the SE based RBF, which is defined as,
\begin{eqnarray} 
\label{eq:rb_func}
S_n(\mathcal{L}_n, \mathcal{L})  = \log_2 \Bigg (1 + \frac{ P^t_n / B_{\max} }{  \gamma \sigma^2  || \mathcal{L} - \mathcal{L}_n ||_2^{\alpha} + \beta \sigma^2 } \Bigg),
\end{eqnarray}
and $C_S(\mathcal{L})$ is the corresponding SE based RBF network. In the following, we provide the existence and uniqueness property of SE based RBF network as summarized below.

\begin{Thm}[Existence of RBF Network] \label{thm:arbitrarily_approx} 
For any continuous traffic distribution $D_T$, and any location $\mathcal{L}_m$ defined on $\mathbb{R}^d$, there exists an SE based RBF network $C_S(\mathcal{L}_m) =  \sum_{n=1}^{N_{BS}} B_n S_n(\mathcal{L}_n, \mathcal{L}_m)$ with coefficients $\{B_{n}\}$, $\{P_{n}^{t}\}$, and $\{\mathcal{L}_n\}$, such at for any $\mathcal{L}_m \in \mathbb{R}^d$,
$$\| C_S(\mathcal{L}_m) - D_T(\mathcal{L}_m) \|_2 \leq \epsilon.$$ 
\end{Thm}
\IEEEproof Please refer to Appendix~\ref{appendix:arbitrarily_approx} for the proof.
\endIEEEproof

\begin{Thm}[Uniqueness of RBF Network] \label{thm:optimal_para}
If the set of all possible RBF networks, e.g., $\mathcal{T}_{N_{BS}} = \{C_S(\mathcal{L}) | C_S(\mathcal{L}) = \sum_{n=1}^{N_{BS}} B_n S_n(\mathcal{L}_n, \mathcal{L}), B_n, P^t_n \in \mathbb{R}^+, \mathcal{L}_n \in \mathbb{R}^d  \}$, is a Chebyshev set \cite{efimov1961approximative}, then there exists an unique RBF network with parameters $\{\Bar{\mathcal{L}_n}\}, \{\Bar{B_n}\}, \{\Bar{P^{t}_n}\}$, such that
\begin{eqnarray}
    \{\Bar{\mathcal{L}_n}\}, \{\Bar{B_n}\}, \{\Bar{P^{t}_n}\} = \underset{\{\mathcal{L}_n\}, \{B_n\},\{P^t_n\}}{\arg\min} \xi\left(C_S, D_{T}\right).
\end{eqnarray}
\end{Thm}
\IEEEproof
Please refer to Appendix~\ref{appendix:optimal_para} for the proof. 
\endIEEEproof

With the existence and uniqueness properties of an SE based RBF network as illustrated in Theorem~\ref{thm:arbitrarily_approx} and~\ref{thm:optimal_para}, we show that the JS divergence, $\xi\left(C_S, D_{T}\right)$, can be minimized by alternatively optimizing the BS location, the bandwidth, and the transmit power, where the detailed mathematical manipulations are summarized in Appendix~\ref{appendix:optimal_power}. Meanwhile, if $C_{Tot} \geq D_{Tot}$ holds true, the optimal value of IREE, $\eta^{\star}_{IREE}$, is bounded by the following lemma.

\begin{Lem}[Optimal IREE Bound] 
\label{lem:bounded_problem} The optimal value of IREE, $\eta^{\star}_{IREE}$, can be bounded by the following expression.
\begin{eqnarray} 
\label{eq:}
\frac{D_{Tot}\left[1 - \xi\left(\Bar{C}_S, D_{T}\right)\right]}{\lambda P_{\max} + N_{BS} P^{c}} \leq \eta^{\star}_{IREE}
\leq \frac{D_{Tot}\left[1 - \xi\left(\Bar{C}_S, D_{T}\right)\right]}{\lambda P^{t}_D + N_{BS} P^{c}},  \nonumber 
\end{eqnarray}
where $\xi\left(\Bar{C}_S, D_{T}\right)$ is the optimal JS divergence as obtained by Theorem~\ref{thm:optimal_para}, and $P^{t}_D $ is the minimum transmit power required to meet the total traffic $D_{Tot}$ given by $P^{t}_D = \frac{B_{\max} \sigma^2 \iint_{\mathcal{A}} L(\mathcal{L}, \mathcal{L}_{\mathcal{A}}) \textrm{d}\mathcal{L}}{V_{\mathcal{A}}}  \left( 2^{\frac{D_{Tot}}{V_{\mathcal{A}} B_{\max}}} - 1 \right)$. $V_{\mathcal{A}}$ and $\mathcal{L}_{\mathcal{A}}$ are the total volume and the central location of area $\mathcal{A}$, respectively.
\end{Lem}
\IEEEproof
Please refer to Appendix~\ref{appendix:bounded_problem} for the proof.
\endIEEEproof

In addition, if we denote $\{\alpha_n>2\}_{n=1}^{N_{BS}}$, $\{\beta_n>0\}_{n=1}^{N_{BS}}$ and $\{\Gamma_n \in \mathbb{S}^d_{++}\}_{n=1}^{N_{BS}}$ to be the path loss exponents and the corresponding normalization factors, we can follow the similar procedures as shown in \cite{poggio1990networks} to construct a linear combination of many SE based RBFs and generalize the above results to more practical path loss models as summarized below.

\begin{Rem}[General Models] 
\label{rem:extension_theorem} 
If the path loss model, $L_n(\mathcal{L})$, follows $L_n(\mathcal{L}) = \left [ (\mathcal{L} - \mathcal{L}_n )^{T} \Gamma_n (\mathcal{L} - \mathcal{L}_n ) \right ]^{\alpha_n/2} + \beta_n$, then Theorem~\ref{thm:arbitrarily_approx} and \ref{thm:optimal_para}, and Lemma~\ref{lem:bounded_problem} still hold true, when the SE based RBFs are given by,
\begin{eqnarray} 
\label{eq:rb_func_path}
S_n^G(\mathcal{L}_n, \mathcal{L}) &=&   \log_2 \Bigg (1 + \frac{ P^t_n }{ B_{\max} \sigma^2 L_n(\mathcal{L})  } \Bigg), \forall n. 
\end{eqnarray}
\end{Rem}

\subsection{Proposed Scheme for IREE Maximization}

\begin{figure*} [t]
\centering
\subfigure[One-shot training strategy.]{
\begin{minipage}[c]{0.45\linewidth}
\centering
\includegraphics[height=7cm,width=8cm]{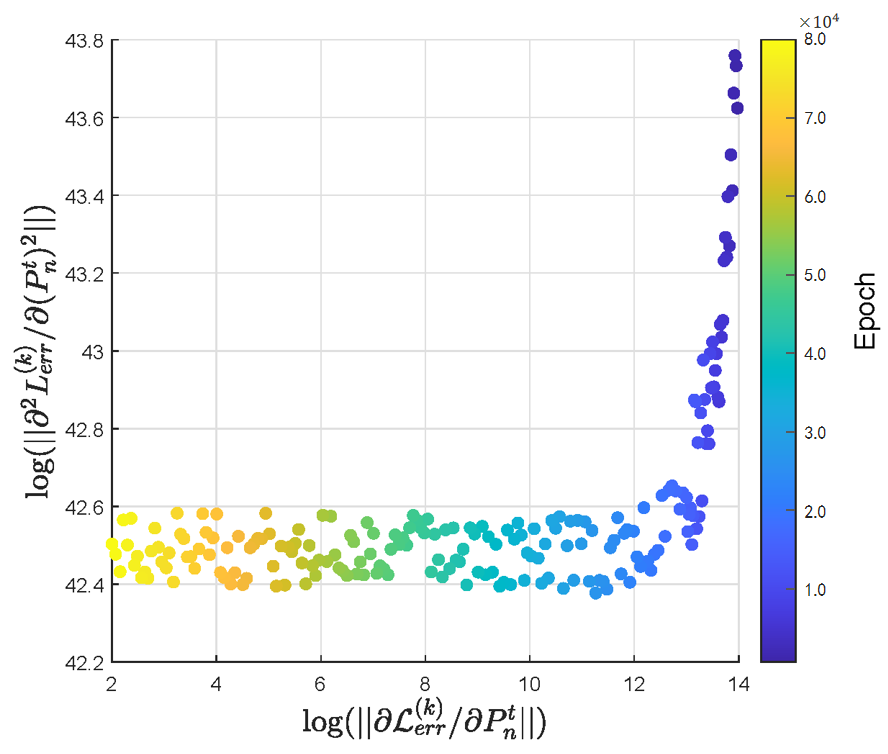}
\label{fig:Exponential_smoothness}
\end{minipage}}
\subfigure[Two-stage training strategy ]{
\begin{minipage}[c]{0.45\linewidth}
\centering
\includegraphics[height=7cm,width=8cm]{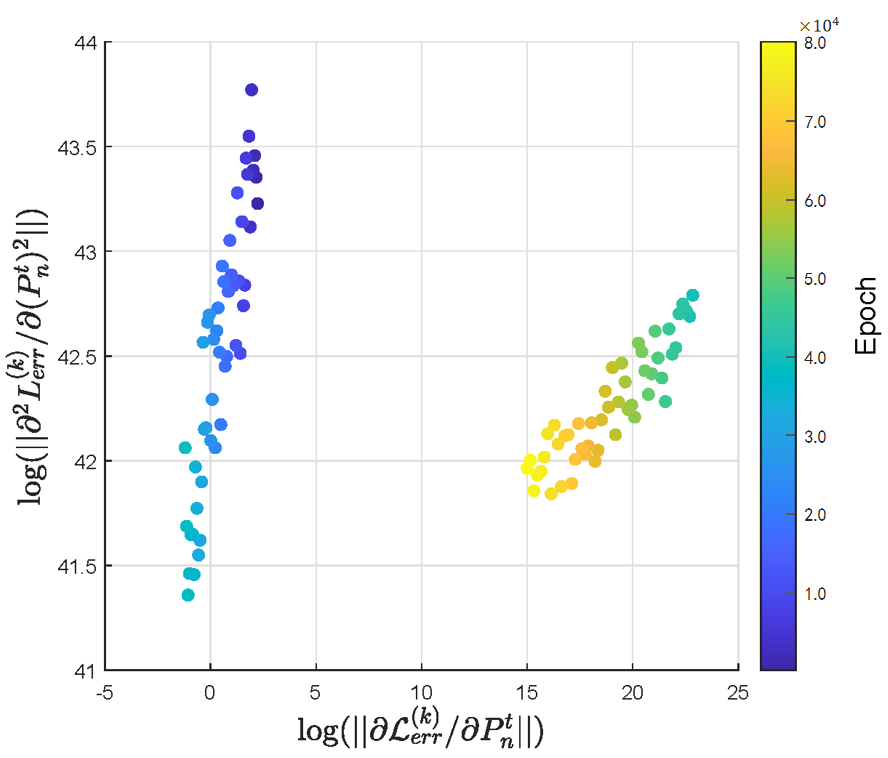}
\label{fig:piecewise_smoothness}
\end{minipage}}

\caption{The norm of second-order gradient versus the norm of first-order gradient on the training trajectory of Adam for proposed RBF network. It can be observed that norm of second-order gradient and the norm of first-order gradient nearly satisfy an exponential relationship using one-shot training strategy while they satisfy a piece-wise linear relationship using the proposed two-stage training strategy. } 
\label{fig:general_smoothness}
\end{figure*}

In order to obtain the optimal IREE, we decouple the original IREE maximization problem into two sub-problems and solve them iteratively as shown in Fig.~\ref{fig:algorithm}. In the first step, we use the aforementioned RBF network to solve Problem~\ref{prob:transformed} for any given $\eta_{IREE}$, while in the second step, we adopt the Dinkelbach's algorithm \cite{dinkelbach1967nonlinear} to update $\eta_{IREE}$ for Problem~\ref{prob:origin}.

\subsubsection{Proposed RBF Network Optimization for Problem~\ref{prob:transformed}} 

For any given IREE in the $k$-th iteration, $\eta_{IREE}^{(k)}$, we can construct the loss function, $L_{err}^{(k)}(\{\mathcal{L}_n\}, \{B_n\},\{P^t_n\} ;\bm{\omega} )$, as,
\begin{eqnarray}
&& L_{err}^{(k)}(\{\mathcal{L}_n\}, \{B_n\},\{P^t_n\};\bm{\omega})  = - \min \bigg\{ \sum_{m=1}^M  C_S(\mathcal{L}_m) , \nonumber \\ 
&& \sum_{m=1}^M  D_T(\mathcal{L}_m) \bigg\}  \times \left[1 - \sum_{m=1}^M \xi\left(C_S(\mathcal{L}_m), D_{T}(\mathcal{L}_m)\right) \right] \nonumber \\
&& + \eta_{IREE}^{(k)}\sum_{n=1}^{N_{BS}}  \Big ( \lambda P^t_n + P^{c} \Big )  + \bm{\omega} \Omega( \{\mathcal{L}_n\}, \{B_n\},\{P^t_n\} ), \nonumber 
\end{eqnarray}
where $\bm{\omega}$ is the  penalty coefficient and $\Omega( \{ \mathcal{L}_n\}, \{B_n\},\{P^t_n\}  ) $ is the penalty term defined by,
\begin{eqnarray}\label{eqn:loss_func_const}
\Omega( \{\mathcal{L}_n\}, \{B_n\},\{P^t_n\} ) = \max \big\{ \zeta_{\min}  - \zeta(\{\mathcal{L}_n\}, \{B_n\},\{P^t_n\}), \nonumber \\
0  \big\}  + 
\max \left \{ \sum_{n=1}^{N_{BS}} B_n - B_{\max}, 0  \right \} + \max \left \{ \sum_{n=1}^{N_{BS}} P^t_n - P_{\max}, 0  \right\}.\nonumber
\end{eqnarray}

By constructing the above loss function, we can design an SE based RBF network with the multi-layer perceptron (MLP) like structure to minimize $L_{err}^{(k)}( \{\mathcal{L}_n\}, \{B_n\},\{P^t_n\} ;\bm{\omega} )$ in each step, which consists of an input layer, a hidden layer, and an output layer as shown in Fig.~\ref{fig:algorithm}.

\begin{itemize}
    \item{\em Input Layer}: The input layer consists of $M$ sampled locations with two dimensional geographic coordinates given by $\{ \mathcal{L}_m \}_{m=1}^M$.
    \item {\em Hidden Layer}: The hidden layer consists of $N_{BS}$ SE based RBF neurons, and each RBF neuron is designed according to \eqref{eq:rb_func}. With different locations and transmit power configured for each RBF neuron, e.g., $\{\mathcal{L}_n\}, \{P^t_n\}$, the hidden layer can output the SE based RBF functions of $M$ sampled locations, e.g., $\{S_n^{G}(\mathcal{L}_n,\mathcal{L}_m)\}_{m=1}^{M}$, for the $n^{th}$ BS (RBF neuron).
    \item {\em Output Layer}: The output layer linearly combines $N_{BS}$ SE based RBF neurons with coefficients given by $\{B_n\}$, and outputs the lower bound 
    $\{C_S(\mathcal{L}_m)\}_{m=1}^M$ as defined in \eqref{eqn:def_Cs}, which eventually minimize the loss function, $L_{err}^{(k)}(\{\mathcal{L}_n\}, \{B_n\},\{P^t_n\};\bm{\omega})$.
\end{itemize}

\begin{table} [t] 
\centering 
\caption{Detailed configurations of proposed RBF network}  
\label{tab:rbf_configurations}
\footnotesize
\begin{tabular}{c | c}  
\toprule
Input Layer & \makecell{Two dimensional geographic \\ coordinates: $\{ \mathcal{L}_m \}_{m=1}^M$ } \\
\midrule
Hidden Layer &  RBF: $\{S_n^{G}(\mathcal{L}_n,\mathcal{L}_m)\}_{m=1}^{M}$ defined in\eqref{eq:rb_func}  \\
\midrule
Output Layer & \makecell{Capacity lower bound:\\  $\{C_S(\mathcal{L}_m)\}_{m=1}^M$  defined in \eqref{eqn:def_Cs}} \\
\midrule 
Loss function  &  $L_{err}^{(k)}(\{\mathcal{L}_n\}, \{B_n\},\{P^t_n\};\bm{\omega})$ \\
\midrule 
Batch size & $M$ (full batch training) \\
\midrule 
\makecell{Training epoch in\\ each iteration}  & $N_{epoch} = 10000$ \\
\midrule 
Neural network optimizer& Adam optimizer \cite{kingma2014adam} \\
\bottomrule
\end{tabular}
\end{table} 

The detailed configuration of proposed RBF network is summarized in Table~\ref{tab:rbf_configurations}. For convenience, we also implement a softmax layer to compare $\{ C_S(\mathcal{L}_m) \}_{m=1}^M$ and $\{ D_T(\mathcal{L}_m) \}_{m=1}^M$ and obtain the JS divergence, $\xi\left(C_S, D_{T}\right)$. 

\subsubsection{Proposed Dinkelbach's Algorithm for Problem~\ref{prob:origin}} 

For any given IREE in the $k$-th iteration, $\eta^{(k)}_{IREE}$, we obtain the optimized parameters $\{\mathcal{L}_n^{(k)}\}, \{B_n^{(k)}\},\{P^{t,(k)}_n\}$ based on the proposed RBF network in the previous subsection, which minimize the loss function, $L_{err}^{(k)}(\{\mathcal{L}_n\}, \{B_n\},\{P^t_n\};\bm{\omega})$. In this subsection, we then update $\eta_{IREE}$ based on Dinkelbach's algorithm, where the updated IREE, $\eta^{(k+1)}_{IREE}$, is given by,
\begin{eqnarray}\label{eqn:iree_update}
\eta^{(k+1)}_{IREE} = \frac{\min\{C_{Tot}^{(k)},D_{Tot}\}}{P_T\left(\{P^{t,(k)}_n\}\right)} \times \left[1 - \xi \left(C_S^{(k)}, D_{T}\right) \right].
\end{eqnarray}
In the above expression, $C_{Tot}^{(k)}$ and $C_S^{(k)}$ are calculated based on the optimized parameters $\{\mathcal{L}_n^{(k)}\}, \{B_n^{(k)}\},\{P^{t,(k)}_n\}$.

By iteratively updating $\eta_{IREE}$ and $\{\mathcal{L}_n\}, \{B_n\},\{P^{t}_n\}$, we are able to calculate the optimized value of IREE and the corresponding strategy, which satisfies \eqref{eq:opt} according to Lemma~\ref{lem:optimal_condition}. The entire IREE maximization scheme has been summarized in Algorithm~\ref{alg:Dinkelbach}.

\begin{figure*} [t]
\centering
\subfigure[ Loss function during the training process. ]{
\begin{minipage}[c]{0.45\linewidth}
\centering
\includegraphics[height=7.5cm,width=8cm]{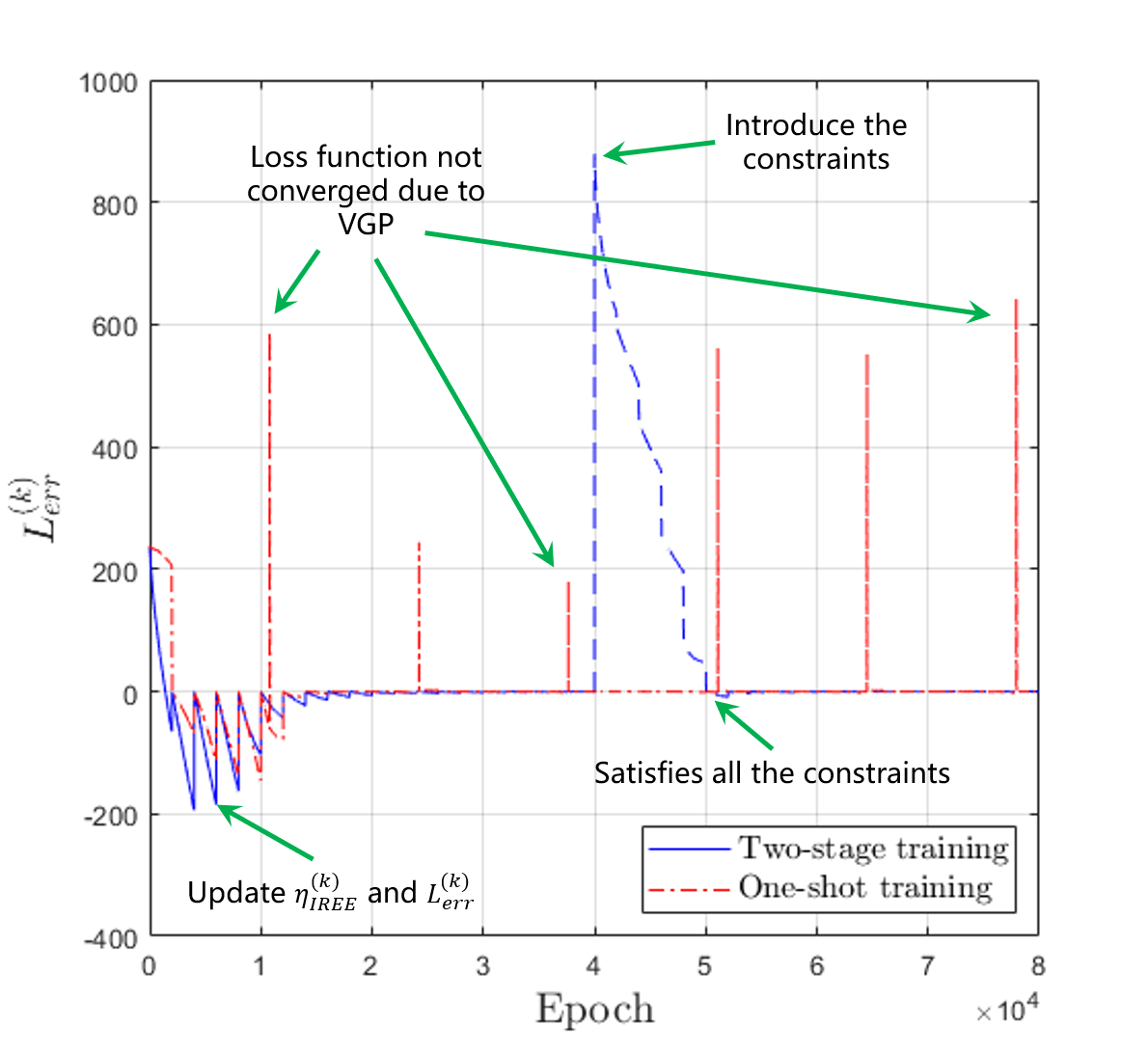}
\label{fig:dynamic_loss}
\end{minipage}}
\subfigure[IREE during the training process. ]{
\begin{minipage}[c]{0.45\linewidth}
\centering
\includegraphics[height=7.5cm,width=8cm]{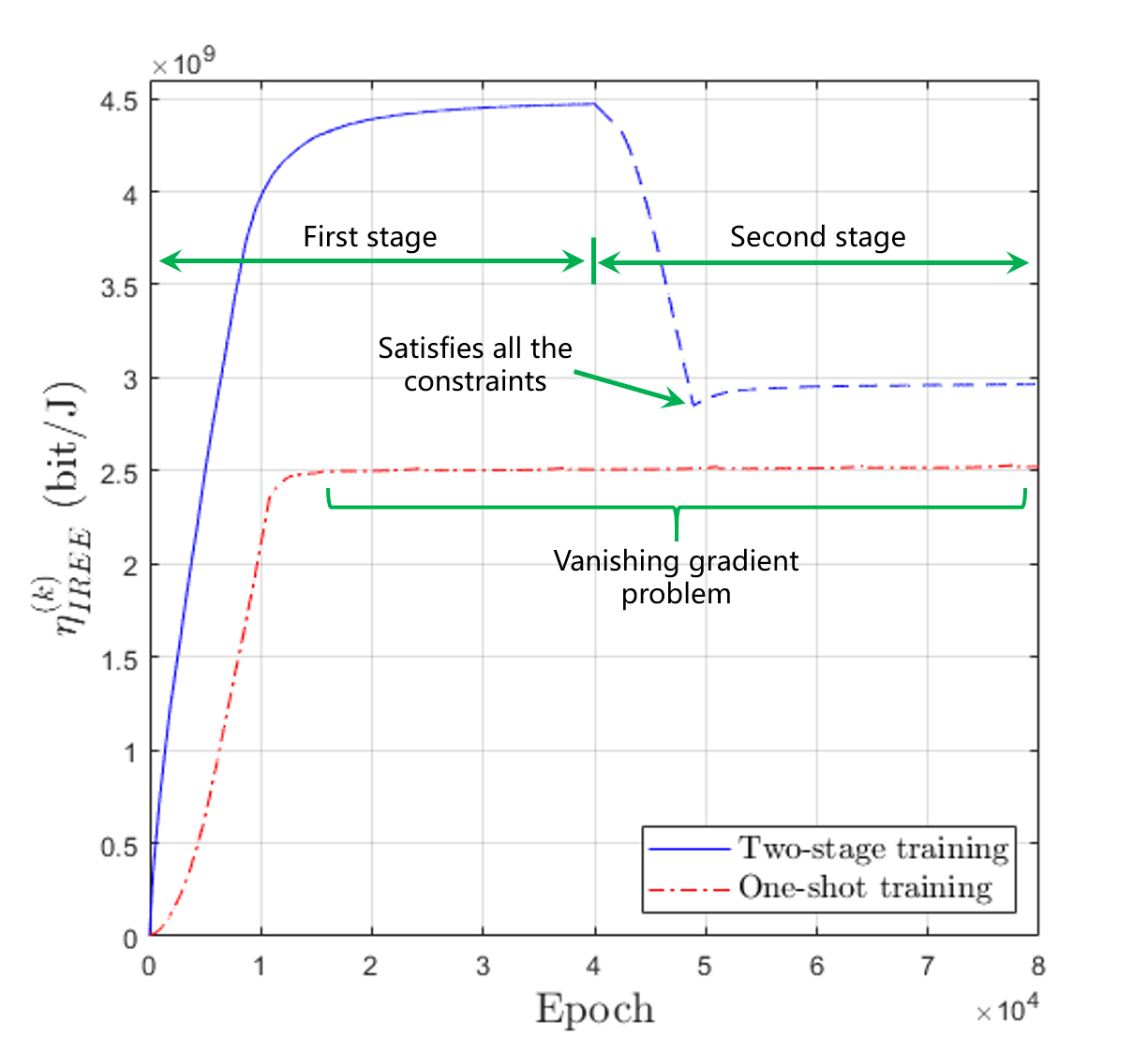}
\label{fig:two_stage_convergence}
\end{minipage}}
\caption{Comparison of loss function/IREE under different training strategies. The traffic and the network configuration is shown in Tab.~\ref{tab:simu_para}. The number of samples $M = 1300$, and the number of epoch in each iteration $N_{epoch} = 2000$. It can be observed that under the one-shot training strategy, the VGP caused by the conflict between the gradient of the objective function and the gradient of the constraints will hinder the further convergence of IREE. While the proposed two-stage training strategy avoids the above problems by processing the gradient of the objective function and constraint conditions in two stages respectively.} 
\label{fig:training_process}
\end{figure*}

\section{Two-stage Training Strategy and Convergence Properties} \label{sect:training} 

In this section, we discuss the famous {\em vanishing gradient problem} (VGP) \cite{hochreiter1998vanishing} when training the proposed RBF network, and introduce a two-stage training strategy accordingly. In addition, we analyze the convergence properties of the entire IREE maximization scheme under this two-stage training strategy as well.

\subsection{Two-stage Training Strategy}
\label{subsec:convergence_challenges}

Although the proposed RBF network contains an MLP like structure, the conventional one-shot training strategy for supervised learning cannot be directly applied as explained below.

\begin{Cla} \label{cla:not_smooth}
Since the loss function for the proposed RBF network, $L_{err}(\{\mathcal{L}_n\}, \{B_n\},\{P^t_n\};\bm{\omega})$, does not satisfy the $(L_0,L_1)$ smoothness \cite{zhang2019gradient}, the conventional one-shot training strategy will result in the VGP.
\end{Cla}
\IEEEproof Please refer to Appendix~\ref{appendix:not_smooth} for the proof. 
\endIEEEproof

To tackle this problem, we propose a two-stage training strategy, where the unconstrained optimization is performed in the first stage and the coefficients are fine-tuned to meet the constraints in the second stage. To be more specific, we set the penalty coefficients $\bm{\omega}$ to zero in the first stage, and optimize the RBF network via standard one-shot supervised learning to achieve a balance between the network utility and the total power consumption. Through this approach, we can get rid of constraints \eqref{constrain:qos}-\eqref{constrain:max_power} and have $ \partial E^{(k)} / \partial P^t_n = 0$, where $E^{(k)}$ is given by $- \min \Big\{ \sum_{m=1}^M  C_S(\mathcal{L}_m), \sum_{m=1}^M  D_T(\mathcal{L}_m) \Big\}  \times \left[1 - \sum_{m=1}^M \xi\left(C_S(\mathcal{L}_m), D_{T}(\mathcal{L}_m)\right) \right] + \eta_{IREE}^{(k)}\sum_{n=1}^{N_{BS}}  \big( \lambda P^t_n + P^{c} \big)$. In the second stage, we choose $\bm{\omega}$ to be sufficiently large, and the partial derivative becomes,
\begin{eqnarray}
\partial L_{err}^{(k)}(\{\mathcal{L}_n\}, \{B_n\},\{P^t_n\};\bm{\omega}) / \partial P^t_n = \partial E^{(k)} / \partial P^t_n + \nonumber \\ 
\bm{\omega} \times \partial \Omega( \{\mathcal{L}_n\}, \{B_n\},\{P^t_n\} ) / \partial P^t_n,
\end{eqnarray}
which is strictly greater than zero if constraints \eqref{constrain:qos}-\eqref{constrain:max_power} are not satisfied, and eventually prevent the VGP. 

In Fig.~\ref{fig:Exponential_smoothness} and Fig.~\ref{fig:piecewise_smoothness}, we plot the $(L_0,L_1)$ smoothness, in terms of the norm of second-order gradient ($||\partial^2 L_{err}^{(k)}(\{\mathcal{L}_n\}, \{B_n\},\{P^t_n\};\bm{\omega}) / \partial (P^t_n)^2||$) versus the norm of first-order gradient ($|| \partial L_{err}^{(k)}(\{\mathcal{L}_n\}, \{B_n\},\{P^t_n\};\bm{\omega}) / \partial P^t_n ||$) relation to show the VGP. As depicted in Fig.~\ref{fig:Exponential_smoothness}, when the conventional one-shot training strategy is adopted, the norm of second-order gradient decreases exponentially with respect to the norm of first-order gradient, which breaks the $(L_0,L_1)$ smoothness condition and results in the VGP. For the proposed two-stage training strategy, it ensures a piece-wise linear relationship between the norms of second-order and first-order gradients, which guarantees the $(L_0,L_1)$ smoothness as shown in Fig.~\ref{fig:piecewise_smoothness}.

\subsection{Convergence Properties and Complexity Analysis } 

By applying the proposed two-stage training strategy, we are able to gradually train the proposed RBF network for any given IREE value, and iteratively obtain the optimized IREE as well. In addition, the convergence properties of the proposed IREE maximization scheme can be guaranteed and summarized as below.

\begin{Thm}[Convergence Properties] \label{thm:conver_analy}
If the loss function $L_{err}^{(k)}(\{\mathcal{L}_n\}, \{B_n\},\{P^t_n\};\bm{\omega}) $ satisfies the $(L_0,L_1)$ smoothness and $ \lim_{k \rightarrow \infty} L_{err}^{(k)}(\{\mathcal{L}_n\}, \{B_n\},\{P^t_n\};\bm{\omega}) = 0$, then $\eta^{(k)}_{IREE}$ converges to the optimal IREE value, i.e., $ \lim_{k \rightarrow \infty} \eta^{(k)}_{IREE} = \eta^{\star}_{IREE}$.
\end{Thm}
\IEEEproof Please refer to Appendix~\ref{appendix:conver_analy} for the proof. 
\endIEEEproof

\begin{figure*} [t]
\centering
\subfigure[Network capacity and traffic distributions under conventional EE maximization scheme \cite{lee2021multiagent}, the JS divergence $\xi\left(C_{T}, D_{T}\right) = 0.30$. ]{
\begin{minipage}[c]{0.45\linewidth}
\centering
\includegraphics[height=8cm,width=8cm]{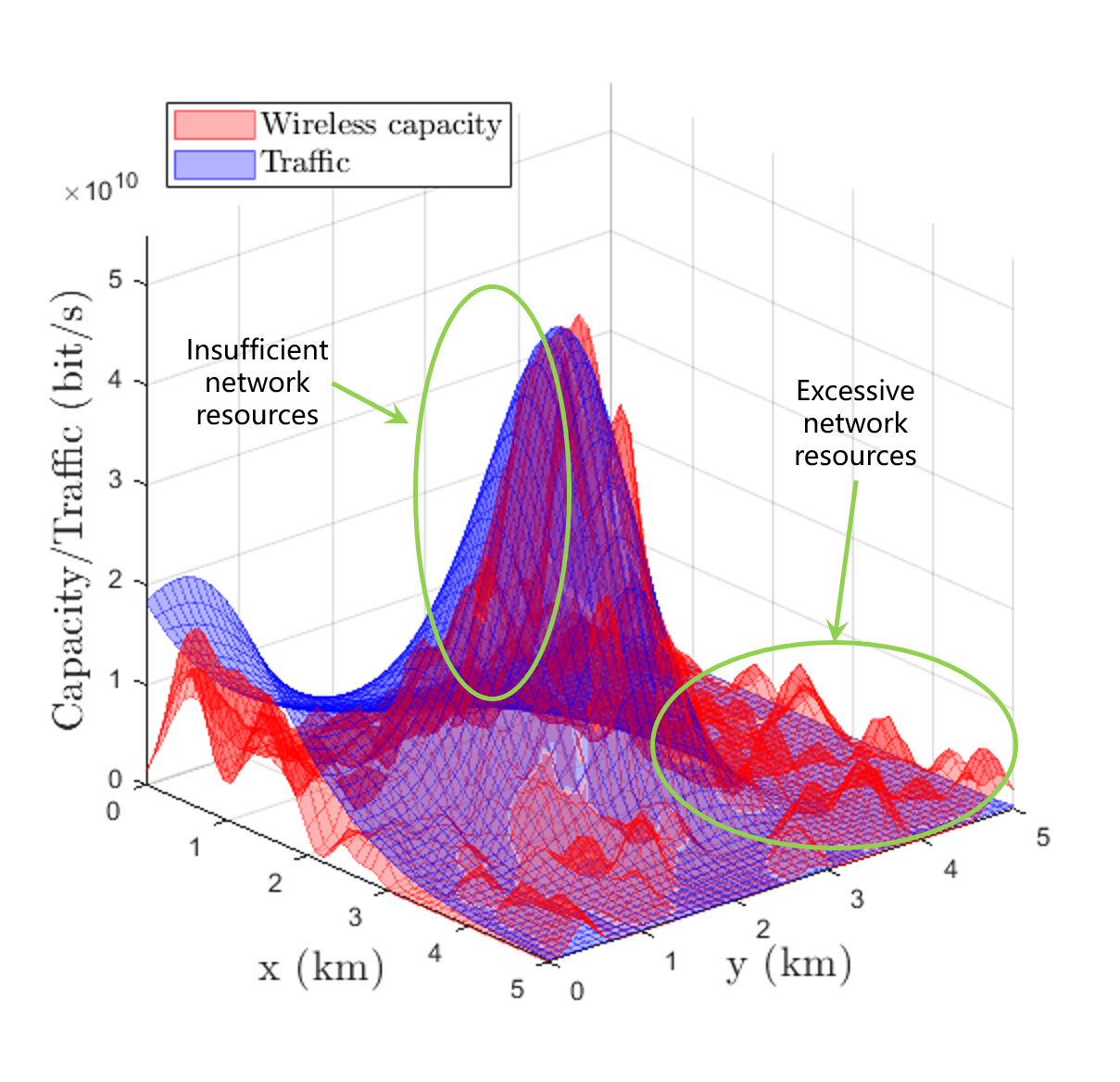}
\label{fig:ee_net_tra_dist}
\end{minipage}}
\subfigure[ Network capacity and traffic distributions under proposed IREE maximization scheme, the JS divergence $\xi\left(C_{T}, D_{T}\right) = 0.05$. ]{
\begin{minipage}[c]{0.45\linewidth}
\centering
\includegraphics[height=8cm,width=8cm]{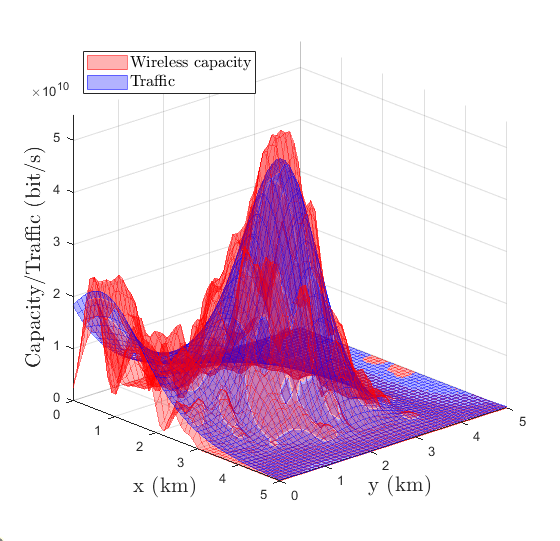}
\label{fig:iree_net_tra_dist}
\end{minipage}}

\caption{ Numerical examples for the traffic and the network capacity distributions under $P_{\max} = 20$ dBW. With the powerful approximation ability of the RBF network, the proposed IREE maximization scheme is able to match the traffic with the network capacity and thus avoids potential resource wastage. } 
\label{fig:net_distri_compare}
\end{figure*}

\begin{table} [t] 
\centering 
\caption{Simulation Parameters}  
\label{tab:simu_para}
\footnotesize
\begin{tabular}{c | c }  
\toprule
Number of BSs, $N_{BS}$ & $400$  \\

\midrule
Height of BSs & $35$ m \\

\midrule
Total Bandwidth, $B_{\max}$ & $36$ GHz \\

\midrule  
Circuit power, $P^{c}$ & $5$ W \\

\midrule  
Efficiency of power amplifier, $1/ \lambda $ & $38\%$ \\

\midrule 
Path loss (dB) & $35 +38\log_{10}(d)$ \cite{6334508} \\

\midrule 
Power spectral density of noise, $\sigma^2$ & $-174$ dBm/Hz \\

\midrule 
User traffic & Log-normal distribution \cite{lee2014spatial} \\

\bottomrule
\end{tabular}  
\end{table}

For illustration purpose, we plot the value of the loss function and the corresponding IREE value versus the index of each epoch in Fig.~\ref{fig:training_process}. As shown in Fig.~\ref{fig:dynamic_loss}, we can observe gradual declining of the loss function during $N_{epoch}$ epochs for both one-shot and two-stage training in the beginning. After $N_{epoch}$ epochs, we update $\eta^{(k)}_{IREE}$ and calculate the corresponding loss function, $L_{err}^{(k)}(\{\mathcal{L}_n\}, \{B_n\},\{P^t_n\};\bm{\omega})$. For one-shot training strategy, it can be observed that the loss function suffers from the VGP and cannot converge within the training period, while for the two-stage training strategy, the loss function converges to zero. This is because the proposed two-stage training strategy guarantees the $(L_0,L_1)$ smoothness as explained before. As shown in Fig.~\ref{fig:two_stage_convergence}, $\eta^{(k)}_{IREE}$ converges to global optima with $\bm{\omega} = 0$ in the first stage, and converges to the local optima with sufficiently large $\bm{\omega}$ in the second stage. Through this approach, we can avoid the VGP and find the optimized IREE finally. 

In the above algorithm development, we replaced $C_T(\mathcal{L})$ with $C_S(\mathcal{L})$ to obtain theoretical results and reduce the implementation complexity, while the optimality gap is summarized in the following lemma. 

\begin{Lem}[Optimality IREE Gap of the Proposed Scheme] 
\label{lem:iree_gap}
Denote $\Bar{\eta}^{S}_{IREE}, \xi^{S}$ and $\Bar{\eta}^{T}_{IREE}, \xi^{T}$ to be the derived optimal IREEs and the corresponding JS divergences using $C_S(\mathcal{L})$ and $C_T(\mathcal{L})$, respectively. The upper bound of this optimality gap $\Delta \eta \triangleq |\Bar{\eta}^{T}_{IREE} - \Bar{\eta}^{S}_{IREE} | $ is given by,
\begin{eqnarray}
\label{eq:delta_eta_bound}
\Delta \eta \leq 
\begin{cases}
& \frac{ (1 - \xi^{S})(C^{T}_{Tot}- C^{S}_{Tot} ) +  \xi^{S,T} C^{T}_{Tot}   }{(1 - \xi^{S} +  \xi^{S,T} ) P_T} , \text{if } C^{S}_{Tot} \leq D_{Tot}, \\
& \frac{ D_{Tot} \xi^{S,T}}{P_T}, \text{if } C^{S}_{Tot} > D_{Tot}, 
\end{cases}
\end{eqnarray}
where $ \xi^{S,T}$ is the JS divergence between $C_T(\mathcal{L})$ and $C_S(\mathcal{L})$. 
\end{Lem}
\IEEEproof Please refer to Appendix~\ref{appendix:iree_gap} for the proof.
\endIEEEproof

In the proposed scheme, we require the order of $O(M N_{BS})$ to compute the forward propagation of the RBF network and the corresponding value of loss function. Then, the gradient calculation and network parameter update of the backward propagation require the computational complexities of orders $O(N_{BS})$ and $O(N_{BS}^2)$ \cite{XU202117}, respectively. Together with the number of iterations in Dinkelbach's algorithm, $N_{ite}$, and the number of epochs, $N_{epoch}$, the overall complexity is given by $O(N_{epoch} N_{ite} ( N_{BS}^2 + M N_{BS}))$.

\section{Numerical Results} \label{sect:num_res}

In this section, we illustrate the advantages of the proposed IREE maximization scheme through some numerical examples. To be more specific, we compare the IREE value of the proposed scheme with conventional SE \cite{lai2016joint} or EE \cite{lee2021multiagent} oriented designs, and characterize the IREE-SE trade-off relation in what follows. By investigating the IREE metric under different traffic requirements, we conclude with some design principles in this part as well.

In the following evaluations, we choose a square area with edge length equal to $5$ kilometers, in which $N_{BS} = 400$ BSs are deployed. The overall throughput requirement is $8.9 \times 10^{12}$ bit/s, which follows a standard log-normal distribution \cite{lee2014spatial} with location, scale, and max spatial spread given by $19$, $2.8$, and $0.0012$, respectively. The total bandwidth and transmit power budgets are given by $B_{\max} = 36$ GHz and $P_{\max} = 30$ dBW, and the minimum network utility indicator is limited by $\zeta_{\min} = 0.8$. The detailed configurations of the proposed RBF network are listed in Table~\ref{tab:rbf_configurations} and other simulation parameters, unless otherwise specified, are listed in Table~\ref{tab:simu_para}. 

\subsection{IREE Comparison with Baselines}

In order to verify the effectiveness of the proposed scheme, we compared the proposed scheme with the conventional SE \cite{lai2016joint} or  EE \cite{lee2021multiagent} oriented design. In Fig.~\ref{fig:net_distri_compare}, we compare the network distribution under different schemes, and we can find that the proposed scheme can effectively reduce JS divergence, $\xi$, thereby avoiding potential resource wastage. 

\begin{figure}  [t]
\centering  
\includegraphics[height=9cm,width=8.5cm]{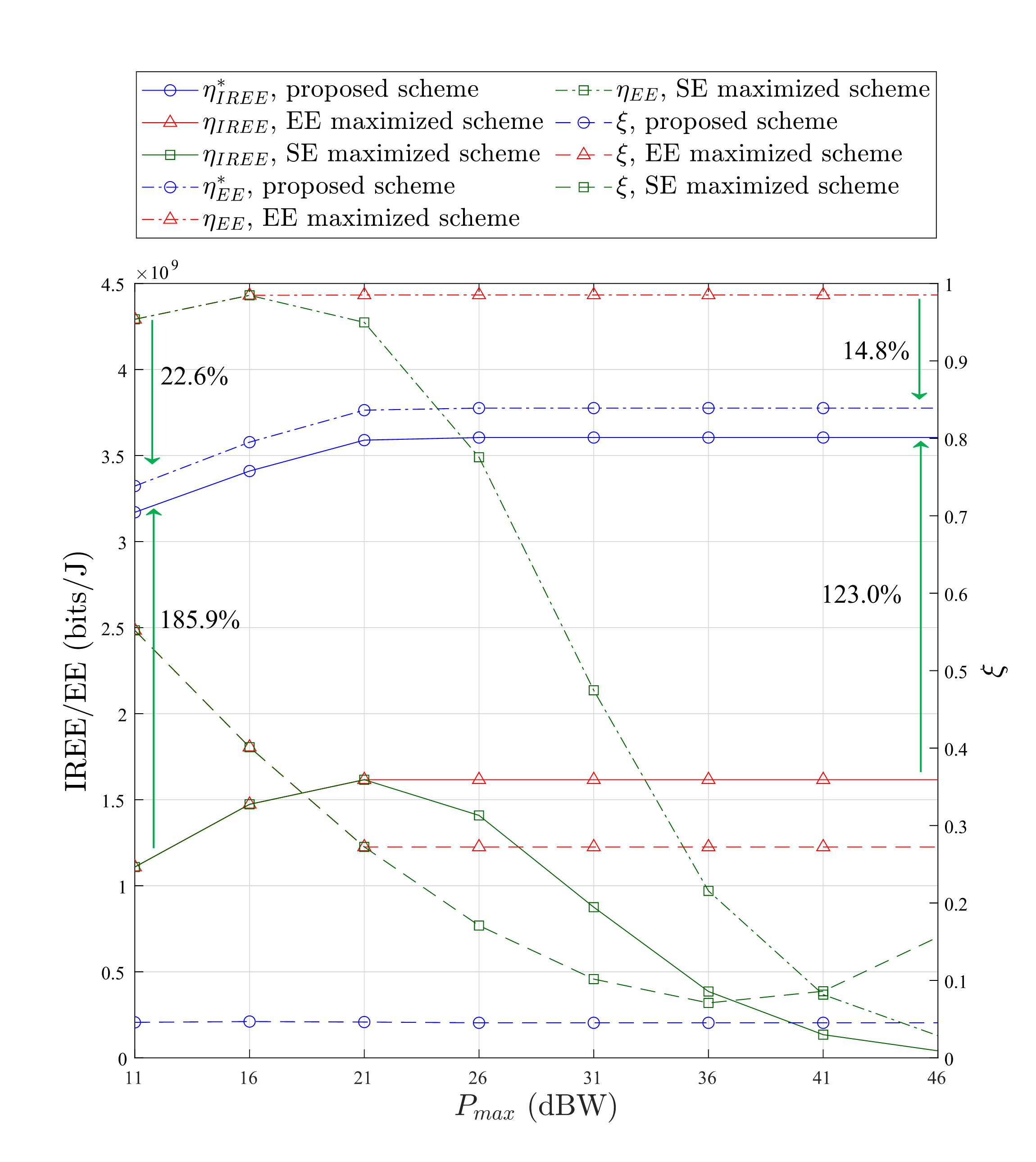}
\caption{The IREE/EE versus $P_{\max}$ relations, and the JS divergence, $\xi$, versus $P_{\max}$ relations for the proposed and other baselines schemes \cite{lai2016joint,lee2021multiagent}. }
\label{fig:iree_ee_compare}
\end{figure}

As illustrated in Fig.~\ref{fig:iree_ee_compare}, the JS divergence of EE oriented design initially decreases and then stabilizes as the transmit power increases, indicating that the maximum EE has been achieved. On the other hand, in the SE oriented design, the JS divergence first decreases and then increases, which can be attributed to the unreasonable spatial distribution of the network resulting from excessive network capacity. However, the proposed scheme is able to consistently maintain a low JS divergence across all power consumption constraints. This means that the IREE maximization scheme outperforms the EE oriented design in terms of maintaining a balanced total network capacity and its distribution, even at low total power consumption. Additionally, Fig.~\ref{fig:iree_ee_compare} highlights that the proposed scheme showcases an earlier convergence of IREE to its maximum value. This can be attributed to the fact that the improvement of utility is no longer enough to pay for the extra power consumption. However, the conventional  EE oriented design is still blindly spending more power to improve the $C_{Tot}$. Compared to the conventional EE oriented design, the proposed scheme achieves a remarkable IREE improvement of $123.0\%$ to $185.9\%$, accompanied by an EE reduction of $14.8\%$ to $22.6\%$. This is because the proposed scheme spends more efforts to reduce the JS divergence rather than contributing to increase the EE metric as explained before.

\begin{figure*} [t]
\centering
\subfigure[Network utility indicator and JS divergence versus the maximum transmit power limit under $B_{\max} = 36$ GHz. ]{
\begin{minipage}[c]{0.45\linewidth}
\centering
\includegraphics[height=8cm,width=8cm]{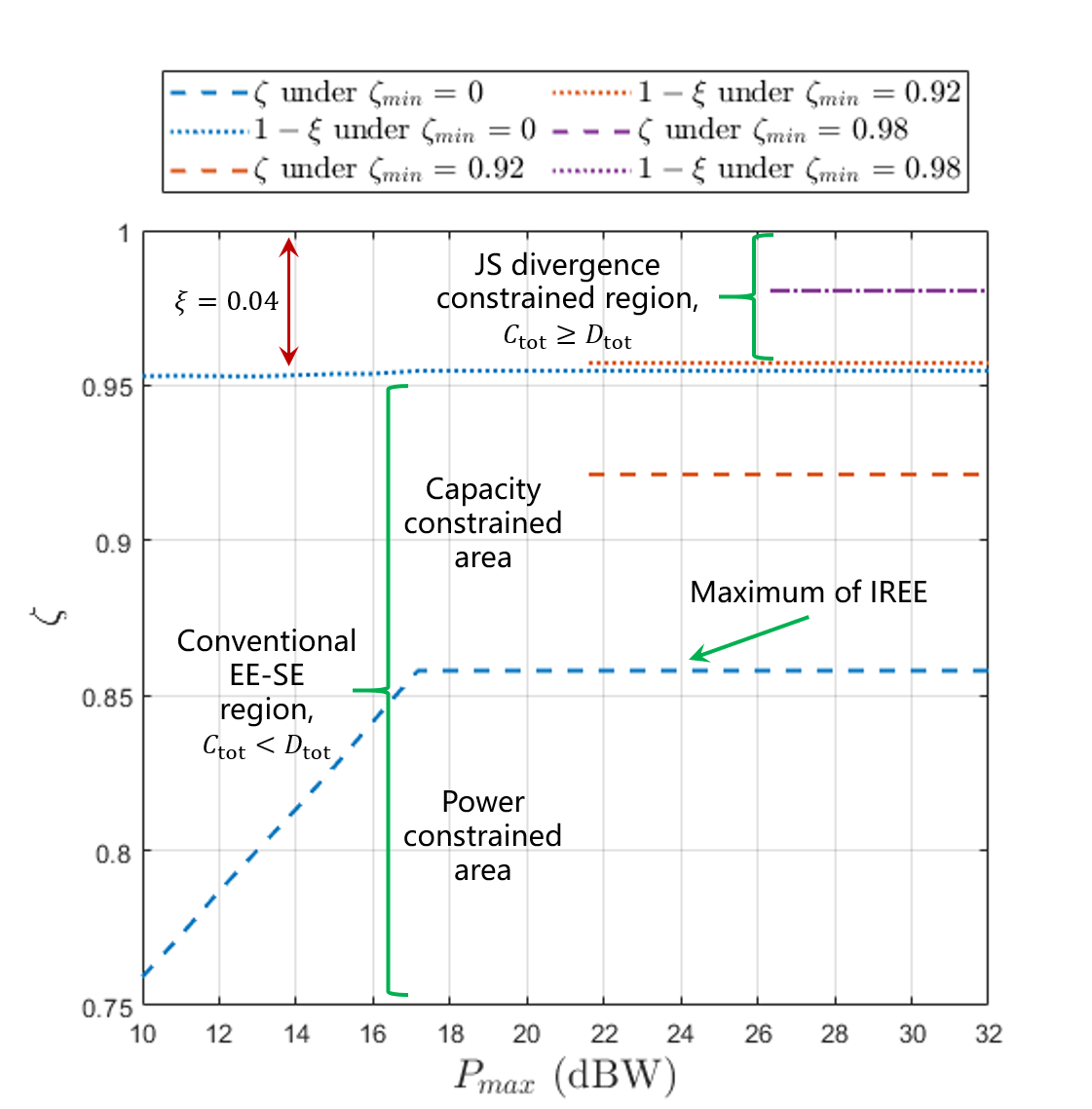}
\label{fig:xi_pmax_1}
\end{minipage}}
\subfigure[ IREE and JS divergence versus SE under different bandwidth limits. ]{
\begin{minipage}[c]{0.45\linewidth}
\centering
\includegraphics[height=8cm,width=8cm]{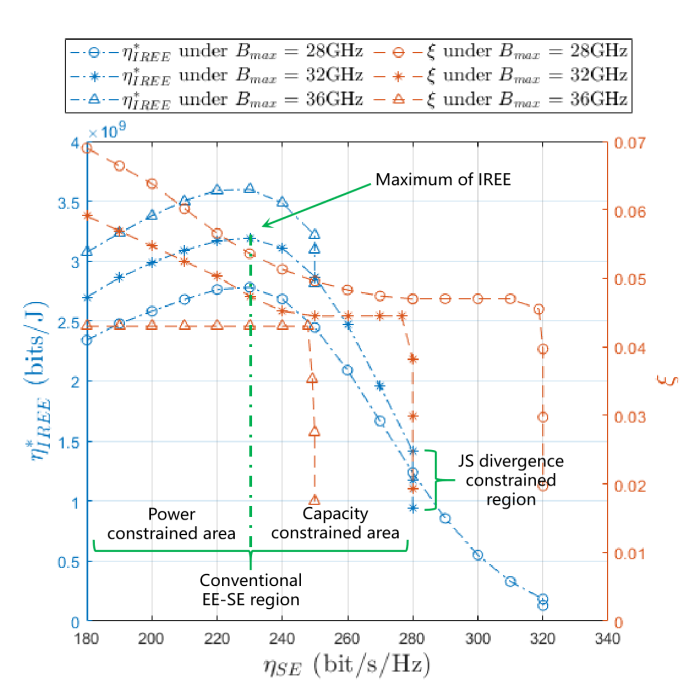}
\label{fig:iree_se_1}
\end{minipage}}
\subfigure[Network utility indicator and JS divergence versus the maximum transmit power limit under $B_{\max} = 40$ GHz.]{
\begin{minipage}[c]{0.45\linewidth}
\centering
\includegraphics[height=8cm,width=8cm]{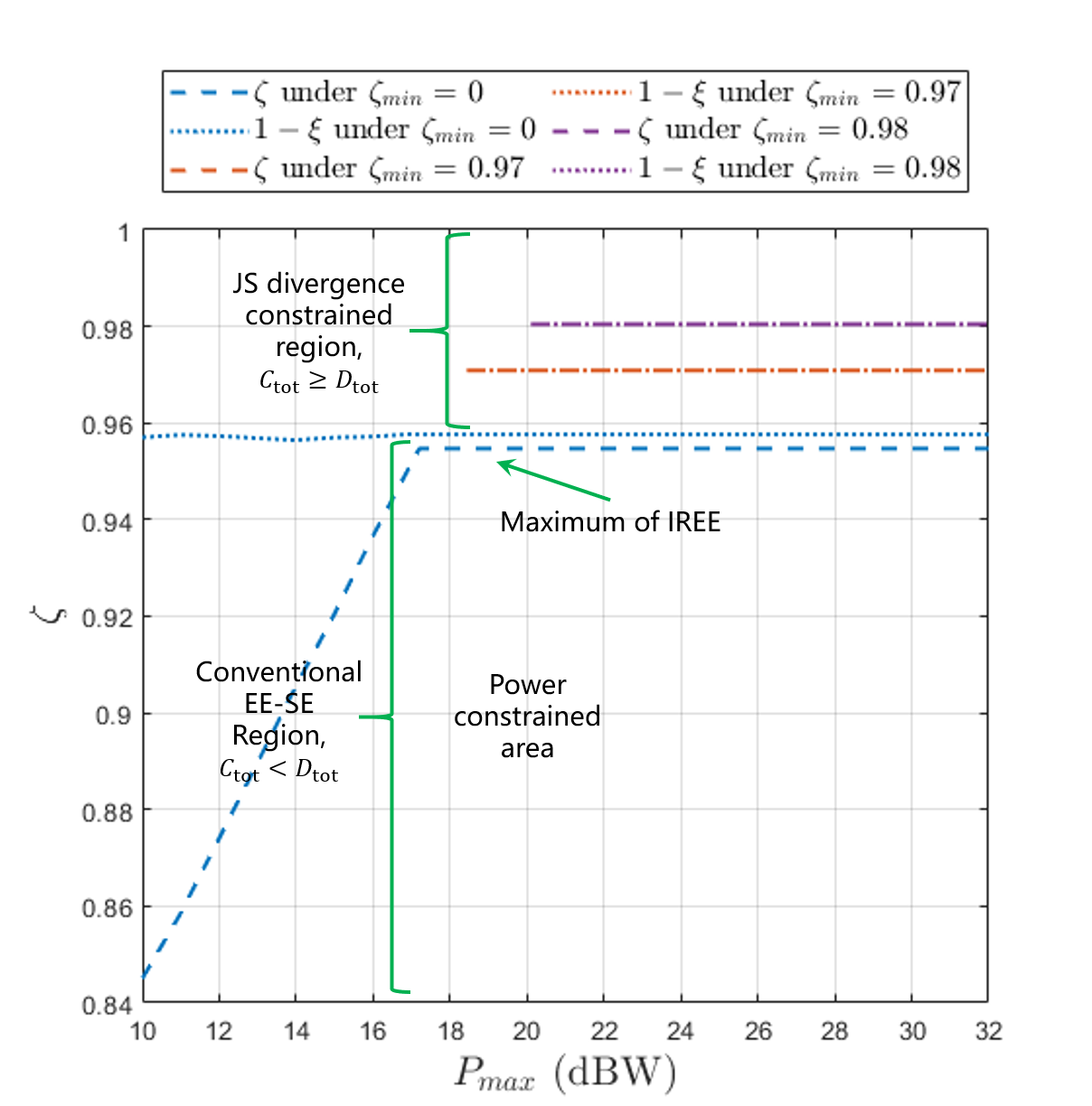}
\label{fig:xi_pmax_2}
\end{minipage}}
\subfigure[ IREE and JS divergence versus SE under different bandwidth limits. ]{
\begin{minipage}[c]{0.45\linewidth}
\centering
\includegraphics[height=8cm,width=8cm]{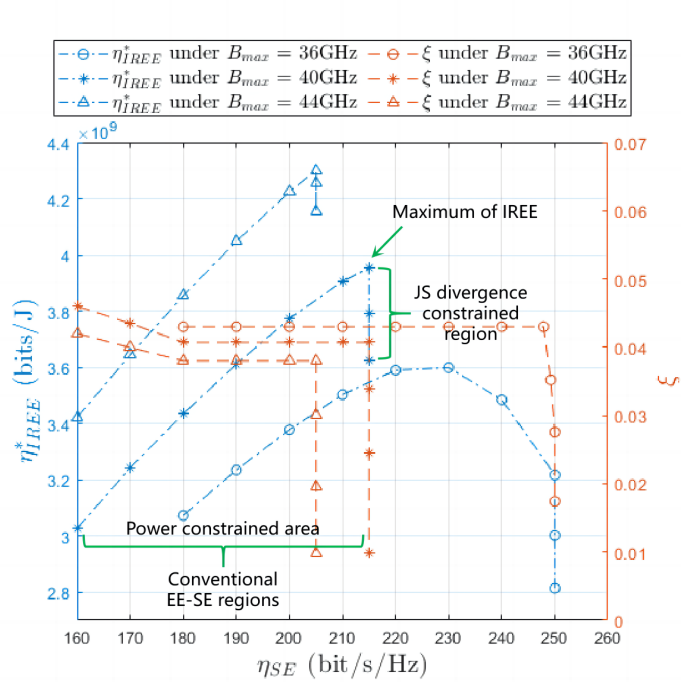}
\label{fig:iree_se_2}
\end{minipage}}
\caption{ Numerical results for the network utility indicator, JS divergence versus maximum transmit power limit and IREE-SE relationships. The amount mismatch and distribution mismatch and their impact on network utility and IREE under different power limits can be observed.} 
\label{fig:xi_pmax_se_iree}
\end{figure*}

\begin{figure}  [t]
\centering
\includegraphics[height=8cm,width=8cm]{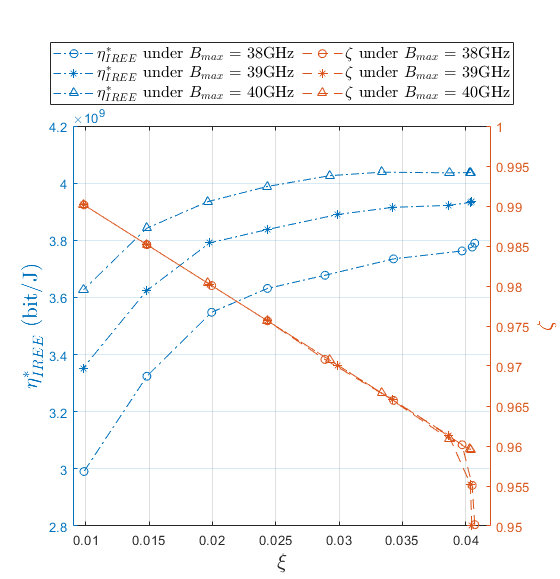}
\caption{IREE versus JS divergence in JS divergence constrained region. IREE performance must be sacrificed to ensure smaller JS divergence, while a linear relationship between the network utility indicator and the JS divergence, $\zeta = 1- \xi$, can be observed.}
\label{fig:iree_xi_tradeoff}
\end{figure}

\subsection{Comparison with Conventional EE-SE Trade-offs}

In Fig.~\ref{fig:xi_pmax_se_iree}, we explore the changes in network utility indicator under different power limit $P_{\max}$ and the corresponding IREE-SE relationships. As shown in Fig.~\ref{fig:xi_pmax_1}, we can divide the network utility indicator $\zeta$ region into two parts including conventional EE-SE region and JS divergence constrained region based on whether the total network capacity $C_{Tot}$ is larger than the total traffic $D_{Tot}$.

As given in Fig.~\ref{fig:iree_se_1} and Fig.~\ref{fig:iree_se_2}, in conventional EE-SE regions, the IREE-SE relationship behaves similarly to conventional EE-SE relationship \cite{zhang2016fundamental}. Specifically, in power constrained area, $\eta^{\star}_{IREE}$ and $\eta_{SE}$ increase simultaneously. There exists a trade-off between $\eta^{\star}_{IREE}$ and $\eta_{SE}$ in capacity constrained area. The JS divergence $\xi$ generally decreases with SE due to the increasing network capacity. The increase in bandwidth has a two-fold effect. Firstly, it causes the capacity constrained area to shrink, enabling the IREE-SE curve to enter the JS divergence constrained region earlier in the process. Eventually, only the power constrained area and the JS divergence constrained region remain, as depicted in Fig.~\ref{fig:xi_pmax_2} and Fig.~\ref{fig:iree_se_2}. Secondly, the enhanced network capacity resulting from the increased bandwidth leads to a decrease in all JS divergences across all SE.

In the JS divergence constrained region, the total network capacity is larger than the total traffic, i.e. $C_{Tot} \geq D_{Tot}$, and hence the network utility is only constrained by the JS divergence as illustrated in Fig.~\ref{fig:xi_pmax_1} and Fig.~\ref{fig:xi_pmax_2}. The SE remains the same since both the total capacity and total bandwidth are unchanged. The extra transmit power introduced at this time will only be used to reduce JS divergence, which leads to a decrease in IREE as shown in Fig.~\ref{fig:iree_xi_tradeoff}. A linear relationship between the network utility indicator and the JS divergence, $\zeta = 1- \xi$, can also be observed in the JS divergence constrained region. 

\subsection{Design Principles}
\label{subsec:design_prin}

\begin{figure}  [t]
\centering  
\includegraphics[height=7cm,width=8cm]{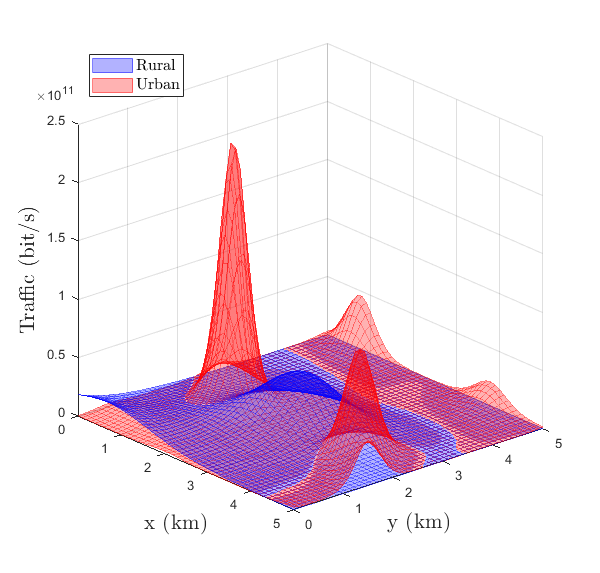}
\caption{Traffics under urban/rural scenario generated from the log-normal distributions \cite{lee2014spatial}. The urban traffic exhibits more heterogeneous spatial variations compared to the rural traffic. }
\label{fig:distri_compare}
\end{figure}

In order to explore the design principles of green networks in different scenarios, we examined both urban and rural traffics. The traffic configurations in the rural scenario are consistent with the configurations above. While in the urban scenario, the total traffic reaches $9.7 \times 10^{12}$ bit/s, and the location, scale, and max spatial spread are given by $19$, $2.4$, and $0.003$, respectively.  As shown in Fig.~\ref{fig:distri_compare}, the spatial variations in urban traffic are significantly more heterogeneous than those in rural traffic, primarily attributed to the dense concentration of people in offices, cafes, and other bustling locations. Another key difference between urban and rural environments lies in the shadowing effects, which can be formulated as a zero-mean log-normal distribution with standard deviation $\chi$ \cite{blaszczyszyn2012quality}. In the following evaluation, we choose $\chi = 10 \, \text{dB}$ for urban scenarios and $\chi = 4 \, \text{dB}$ for rural environments.

\begin{figure*} [t]
\centering
\subfigure[Rural traffic]{
\begin{minipage}[c]{0.45\linewidth}
\centering
\includegraphics[height=8cm,width=8cm]{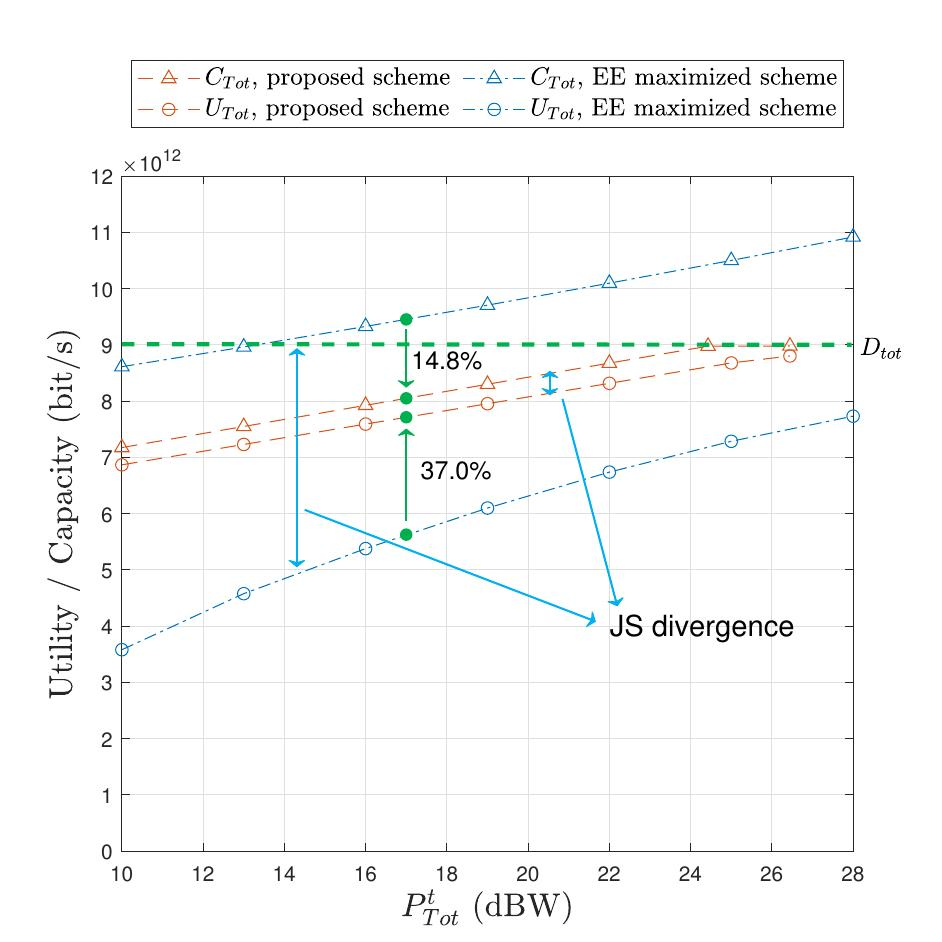}
\label{fig:utility_power_rural}
\end{minipage}}
\subfigure[Urban traffic]{
\begin{minipage}[c]{0.45\linewidth}
\centering
\includegraphics[height=8cm,width=8cm]{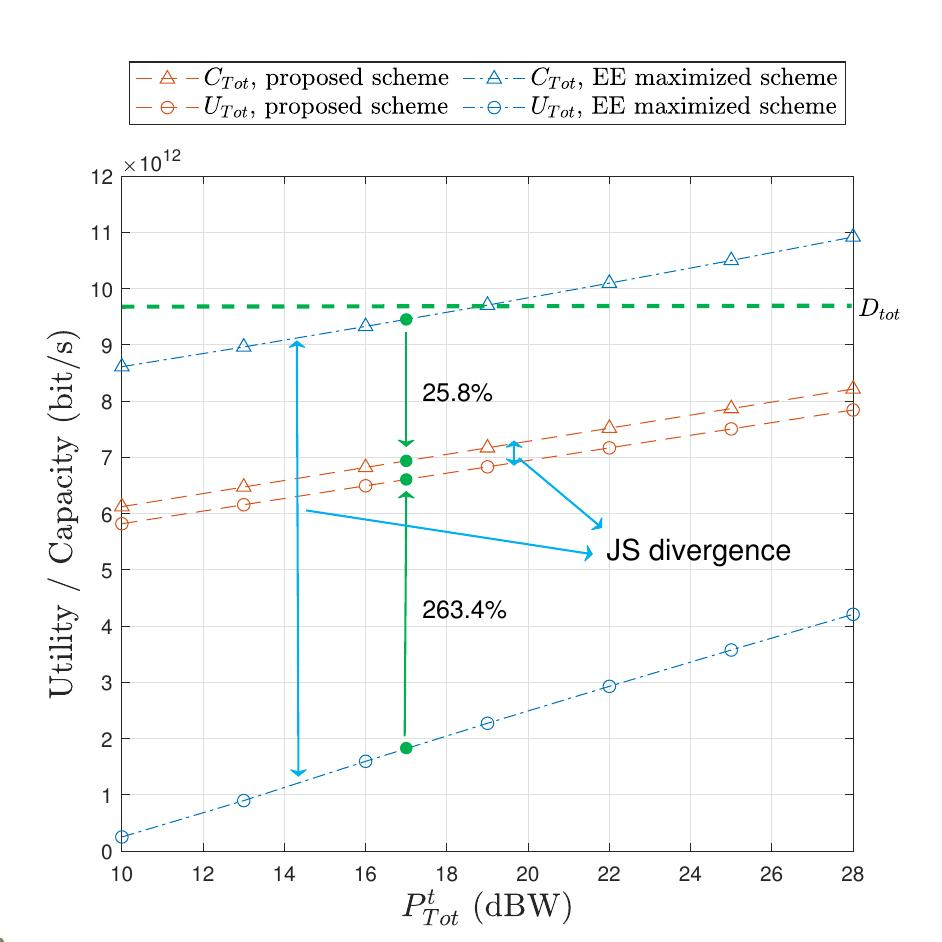}
\label{fig:utility_power_urban}
\end{minipage}}
\caption{The total utility versus the total transmit power relations for the proposed IREE maximization scheme and the EE maximized scheme under rural and urban environment. It can be observed that the proposed IREE maximization solution can effectively enhance the network utility by reducing the JS divergence $\xi$.} 
\label{fig:utility_power_traffic}
\end{figure*}

In order to explore the impact of traffic heterogeneity on the IREE-SE curve, in Fig.~\ref{fig:utility_power_traffic}, we depict the variations of the total network utility $U_{Tot} = \min\{C_{Tot},D_{Tot}\}\left[1 - \xi\left(C_{T}, D_{T}\right)\right]$, the total network capacity $C_{Tot}$ as the total transmit power $P^t_{Tot} = \sum_{n=1}^{N_{BS}} P^t_n$ changes under the above two traffics. Notably, as the spatial variation of traffic becomes significant, the JS divergence experiences a sharp increase. Despite the traditional EE oriented design ensuring a higher total capacity, it fails to guarantee a higher network utility due to the presence of the JS divergence brought by the traffic heterogeneity. This limitation becomes more pronounced in urban scenarios characterized by greater traffic heterogeneity. However, the proposed scheme allocates additional resources to shape the network capacity distribution, prioritizing it over the pursuit of maximum total capacity when significant spatial variations occur. As a result, it effectively enhances network utility by achieving a balance between improving the total capacity and shaping the network capacity distribution.

\begin{figure}  [t]
\centering
\includegraphics[height=9cm,width=8cm]{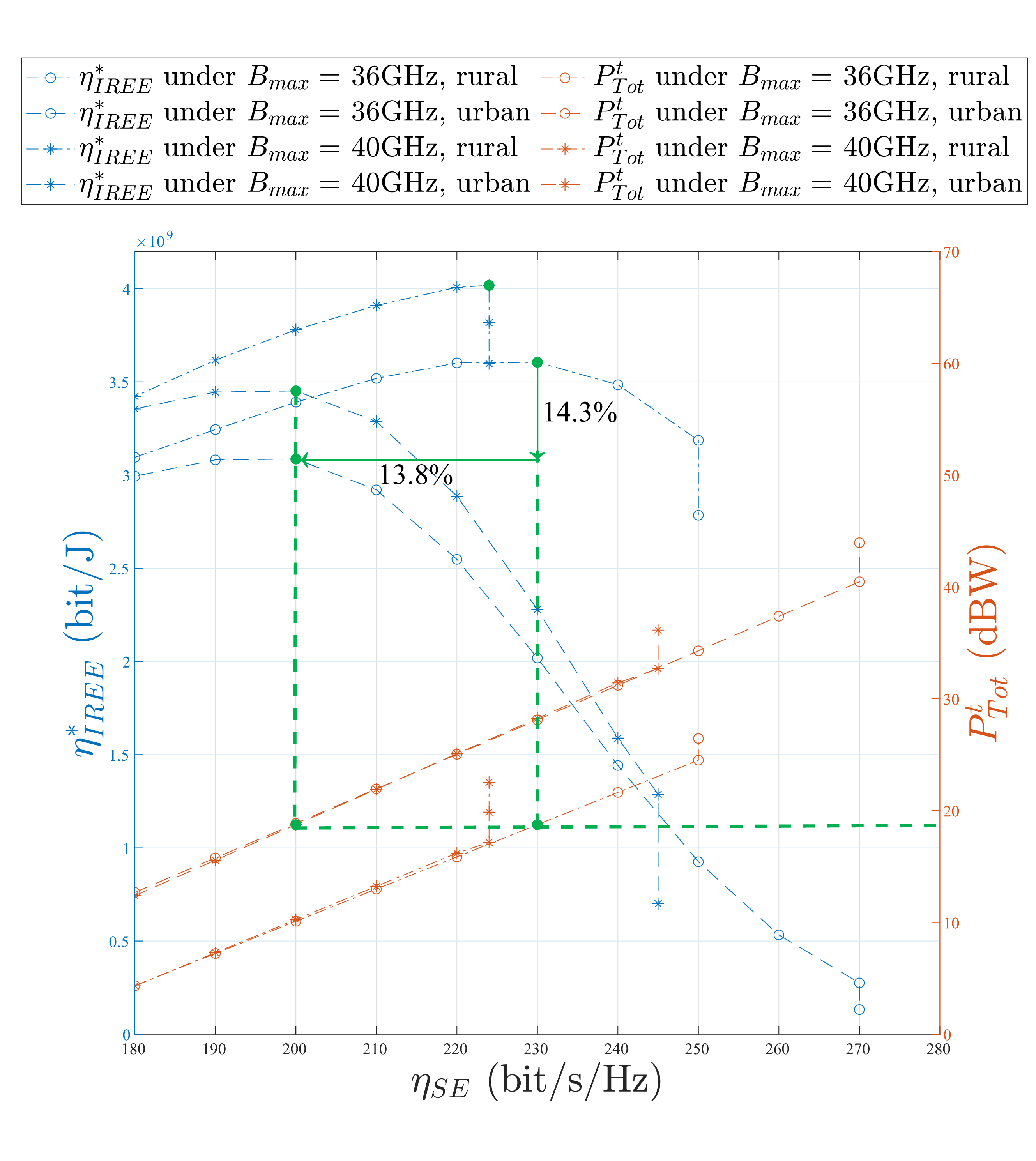}
\caption{The IREE versus SE trade-offs, and the total transmit power versus SE relations under urban/rural environment. }
\label{fig:iree_se_traffic}
\end{figure}

The IREE-SE curve, as depicted in Fig.~\ref{fig:iree_se_traffic}, exhibits a consistent shift towards the lower left direction in urban scenarios. Specifically, the maximum value of IREE decreases by $14.3\%$, accompanied by a corresponding decrease of $13.8\%$ in SE. This shift can be attributed to the effect of the proposed scheme spending more resources to shape the network capacity distribution other than improve the total network capacity under significant spatial variations, as mentioned earlier. The IREE-SE relationship in urban traffic demonstrates a larger capacity constrained area compared to rural traffic. This implies that in urban scenarios, a higher $\eta_{SE}$ can be achieved at the cost of a lower $\eta^{\star}_{IREE}$. Conversely, in rural scenarios, $\eta_{SE}$ is limited by $D_{tot}$ and cannot be further improved.

\section{Conclusion} \label{sect:conc}

In this paper, we present a novel RBF based neural network approach for IREE maximization. By jointly utilizing the SE based RBF network and the Dinkelbach's algorithm, the proposed IREE maximization framework is able to iteratively find the optimized value of IREE and the corresponding resource allocation strategy. The training strategy and the convergence properties of the proposed framework are then analyzed, and the corresponding two-stage training strategy is proposed to ensure the convergence of the proposed RBF network. Through some numerical experiments, we show that the proposed IREE maximization framework outperforms many existing SE or EE oriented designs. Different from conventional EE-SE trade-offs, the IREE-SE trade-offs introduces the new JS divergence constrained region, where the IREE varies and the SE remains unchanged. By studying IREE-SE trade-offs under different traffic requirements, we conclude that the network operators shall spend some efforts to balance the traffic demand and the network capacity distribution in order to improve the IREE performance, especially when the spatial variations of the traffic distribution are significant.

\section*{Acknowledgement}
The authors would like to thank Prof. Shugong Xu from Shanghai University, Prof. Jiandong Li and Prof. Junyu Liu from Xidian University, Prof. Sihai Zhang from University of Science and Technology of China, and anonymous reviewers for their valuable comments, which greatly improve the quality of this paper.

\begin{appendices}

\section{Proof of Theorem~\ref{thm:arbitrarily_approx}}
\label{appendix:arbitrarily_approx}

Considering the problem of recovering a function $C_S$ from a set of data $D_T(\mathcal{L}_m)$, we aim to minimize the error $\sum_{m=1}^{N_{BS}} \left[ C_S(\mathcal{L}_m) - D_T(\mathcal{L}_m) \right ]^2$. This problem is inherently ill-posed due to the infinite number of possible solutions. In order to achieve a specific solution, a common assumption is that the function exhibits smoothness, meaning that similar inputs yield similar outputs. Therefore, the solution to the approximation problem can be obtained by minimizing the following functional \cite{girosi1993priors}:
\begin{eqnarray}
H(C_S) = \sum_{m=1}^{N_{BS}} \left[ C_S(\mathcal{L}_m) - D_T(\mathcal{L}_m) \right ]^2 + \mu \phi(C_S),
\label{functional: approx}
\end{eqnarray}
where $\mu$ is a positive regularization parameter. $\phi(C_S)$ is a smoothness functional given by $\phi(C_S) \triangleq \int_{\mathbb{R}^d} \frac{| \widetilde{ C_S}(s)  |^2}{\widetilde{ G }(s) } \textrm{d}s$, where $\widetilde{}$ indicates the Fourier transform, $\widetilde{ G }$ is some positive function that falls off to zero as $||s|| \rightarrow \infty $ for which the class of functions such that this expression is well defined is not empty.

To make functional $\phi(C_S)$ is a semi-norm so that the associated variational problem \eqref{functional: approx} is well defined, \cite{girosi1993priors} points out that the class of admissible radial basis functions $G$ is the class of conditionally positive definite functions of any order. To this end, we define the function $s(x) \triangleq \log_2 \Bigg (1 + \frac{ 1 }{ v_n x^{\alpha/2} + \kappa_n } \Bigg)$, where $v_n = B_{\max} \gamma \sigma^2 / P^t_n > 0$, $\kappa_n(P^t_n) = B_{\max} \beta \sigma^2 / P^t_n >0$ and $\alpha >2$. Since $s(x)$ is completely monotonic, we have positive definite function $S_n(\mathcal{L}_n, \mathcal{L})$ on any Hilbert space \cite{schoenberg1938metric} defined as 
\begin{eqnarray} 
\label{eq:}
&& S_n(\mathcal{L}_n, \mathcal{L})  = s(||\mathcal{L} - \mathcal{L}_n||^2) \nonumber \\
&& = \log_2 \Bigg (1 + 
\frac{ P^t_n }{ B_{\max} \gamma \sigma^2  || \mathcal{L} - \mathcal{L}_n ||_2^{\alpha} + B_{\max} \beta \sigma^2 } \Bigg). \nonumber \\
\end{eqnarray}

Hence, according to \cite{girosi1993priors}, the solution to the variational problem \eqref{functional: approx} with $G = S_n(\mathcal{L}_m,\mathcal{L}_n)$ is given by $C_S(\mathcal{L}_m) = \sum_{n=1}^{N_{BS}} B_n S_n(\mathcal{L}_n, \mathcal{L}_m)$.

Since $\kappa_n > 0$, $S_n(\mathcal{L}_n, \mathcal{L}_m)$ does not have singularities when $||\mathcal{L}_n-\mathcal{L}_m|| = 0$. Hence, recall Proposition~C.1 in \cite{girosi1990networks}, we have that for every continuous function $D_T$ defined on $\mathbb{R}^d$ and for every Green's function
$G = S_n(\mathcal{L}_n, \mathcal{L}_m)$, there exists a function $C_S(\mathcal{L}_m) = \sum_{n=1}^{N_{BS}} B_n S_n(\mathcal{L}_n, \mathcal{L}_m)$, such that for all $\mathcal{L}_m$ and any positive $\epsilon$ the following inequality holds,
\begin{eqnarray} 
\label{eq:}
|| C_S(\mathcal{L}_m) - D_T(\mathcal{L}_m) ||_2 < \epsilon,
\end{eqnarray}
and hence complete the proof.

\section{Proof of Theorem~\ref{thm:optimal_para}}
\label{appendix:optimal_para}
Considering a Chebyshev set $\mathcal{T}_{N_{BS}} = \{C_S| C_S = \sum_{n=1}^{N_{BS}} B_n S_n(\mathcal{L}_n, \mathcal{L}), B_n, P^t_n \in \mathbb{R}^+, \mathcal{L}_n \in \mathbb{R}^d  \}$, according to \cite{girosi1990networks}, there exists one and only one function $\Bar{C}_S \in \mathcal{T}_{N_{BS}}$, such that for any continuous traffic distribution $D_T$ we have $W \left (\frac{\Bar{C}_S}{C_{Tot}}, \frac{D_{T}}{D_{Tot}} \right )
= \inf_{C_S \in \mathcal{T}_{N_{BS}}}  W \left ( \frac{C_S}{C_{Tot}}, \frac{D_{T}}{D_{Tot}} \right )$, where $W (\cdot, \cdot )$ is the Wasserstein distance \cite{vallender1974calculation}. According to \cite{kolouri2018sliced}, if the network distribution and the traffic distribution have overlapping area, the JS divergence is positively related to Wasserstein distance. Since the $C_S$ is a continuous function in the domain of definition, we have $\xi\left(\Bar{C}_S , D_{T}\right) = \inf_{C_S \in \mathcal{T}_{N_{BS}}} \xi\left(C_S, D_{T}\right) = \min_{C_S \in \mathcal{T}_{N_{BS}}} \xi\left(C_S, D_{T}\right)$.

If the network distribution and the traffic distribution do not have overlapping area, we have $\xi\left(\Bar{C}_S , D_{T}\right) = \min_{C_S \in \mathcal{T}_{N_{BS}}} \xi\left(C_S, D_{T}\right) = 1$. Note  $\{\Bar{\mathcal{L}_n}\}, \{\Bar{B_n}\}, \{\Bar{P^{t}_n}\}$ as the configurations of $\Bar{C}_S$, we have completed the proof.

\section{Derivation of alternative optimization for IREE} 
\label{appendix:optimal_power}

According to Proposition~4.3 in \cite{girosi1990networks} and Theorem~\ref{thm:optimal_para}, with the given transmit power and the location, there exists an optimal bandwidth configuration.

Considering the hypothesis space $\mathcal{T}^{N_{BS}}_P = \{C_S| C_S = \sum_{n=1}^{N_{BS}} B_n S_n(\mathcal{L}_n, \mathcal{L}), P^t_n \in \mathbb{R}^+ \}$, we now prove that
$\mathcal{T}^{N_{BS}}_P$ is a Chebyshev set. Note that $D_T$ is in the space of the continuous functions $\mathcal{C}(U)$ defined on some subset $U$ of $\mathbb{R}^d$. Let $\widehat{C}_S $ be an arbitrary point of $\mathcal{T}^{N_{BS}}_P$, we are looking for the closest point to $D_T$  in $\mathcal{T}^{N_{BS}}_P$, which has to lie in the set
\begin{eqnarray}
\widehat{\mathcal{T}}^{N_{BS}}_P = \{ C_S \in \mathcal{T}^{N_{BS}}_P \big | || C_S - D_T ||_f \leq || \widehat{C}_S  - D_T ||_f \}.
\label{functional: }
\end{eqnarray}
This set is clearly bounded. 


Let $\widetilde{\mathcal{T}}^{N_{BS}}_P$ be the closure of $\mathcal{T}^{N_{BS}}_P$ and  $\widetilde{C}_S \in \widetilde{\mathcal{T}}^{N_{BS}}_P $. Then, there is a sequence $\{C_S^{(k)}\}_{k\geq0}  \subset \mathcal{T}^{N_{BS}}_P$ such that: $C_S^{(k)} \rightarrow \widetilde{C}_S $. As $\{C_S^{(k)}\}_{k\geq0}  \subset \mathcal{T}^{N_{BS}}_P: C_S^{(k)} = \sum_{n=1}^{N_{BS}} B_n S_n(\mathcal{L}_n, \mathcal{L}), P^t_n \in \mathbb{R}^+ $. This gives 
\begin{eqnarray}
\label{eq:}
&\widetilde{C}_S & = \lim_{k\rightarrow \infty} C_S^{(k)} \nonumber \\
&=& \lim_{k\rightarrow \infty} \sum_{n=1}^{{ N_{BS}}} B_n \log_2 \Bigg (1 + 
\frac{ P^{t,k}_n/ B_{\max}  }{\gamma \sigma^2  || \mathcal{L} - \mathcal{L}_n ||_2^{\alpha} + \beta \sigma^2 } \Bigg) \nonumber \\
&=& \sum_{n=1}^{{N_{BS}}} B_n \log_2 \Bigg (1 + 
\frac{ \lim_{k\rightarrow \infty} P^{t,k}_n / B_{\max}  }{ \gamma \sigma^2  || \mathcal{L} - \mathcal{L}_n ||_2^{\alpha} + \beta \sigma^2 } \Bigg). 
\end{eqnarray}
Clearly $\widetilde{C}_S \in \mathcal{T}^{N_{BS}}_P $. This shows $\widetilde{\mathcal{T}}^{N_{BS}}_P \subset \mathcal{T}^{N_{BS}}_P$, hence $\mathcal{T}^{N_{BS}}_P$ is a closed. Let $\mathcal{U} =  \{ a \in \mathcal{C}(U) \big | ||a - f || \leq ||a_0 - f || \}$, this set is clearly closed. Hence $\widehat{\mathcal{T}}^{N_{BS}}_P = \mathcal{U} \cap \mathcal{T}^{N_{BS}}_P$ is closed. Since $\widehat{\mathcal{T}}^{N_{BS}}_P$ is a closed, bounded, finite-dimensional set in a metric linear space, it is compact. According to Proposition 4.3 and Theorem~3.1 in \cite{girosi1990networks}, $\mathcal{T}^{N_{BS}}_P$ is a Chebyshev set.

Therefore, with the given bandwidth and the location, there exists an optimal transmit power configuration according to Theorem~\ref{thm:optimal_para}.

Following similar steps, note
$\mathcal{T}^{N_{BS}}_L = \{C_S| C_S = \sum_{n=1}^{N_{BS}} B_n S_n(\mathcal{L}_n, \mathcal{L}), \mathcal{L}_n \in \mathbb{R}^d \}$. Assume that $\widetilde{\mathcal{T}}^{N_{BS}}_L$ is the closure of $\mathcal{T}^{N_{BS}}_L$ and  $\widetilde{C}_S \in \widetilde{\mathcal{T}}^{N_{BS}}_L $, we need to prove that $\widetilde{C}_S \in \mathcal{T}^{N_{BS}}_L$ in order to prove that $\mathcal{T}^{N_{BS}}_L$ is a Chebyshev set.

Since there is a sequence $\{C_S^{(k)}\}_{k\geq0}  \subset \mathcal{T}^{N_{BS}}_L$ such that: $C_S^{(k)} \rightarrow \widetilde{C}_S $. As $\{C_S^{(k)}\}_{k\geq0}  \subset \mathcal{T}^{N_{BS}}_L: C_S^{(k)} = \sum_{n=1}^{N_{BS}} B_n S_n(\mathcal{L}_n, \mathcal{L}), \mathcal{L}_n \in \mathbb{R}^d $. This gives
\begin{eqnarray}
\label{eq:}
&\widetilde{C}_S & = \lim_{k\rightarrow \infty} C_S^{(k)} \nonumber \\
&=& \lim_{k\rightarrow \infty} \sum_{n=1}^{N_{BS}} B_n \log_2 \Bigg (1 + 
\frac{ P^t_n/ B_{\max} }{ \gamma \sigma^2  || \mathcal{L} - \mathcal{L}_n^k ||_2^{\alpha} + \beta \sigma^2 } \Bigg) \nonumber \\
&=& \sum_{n=1}^{N_{BS}} B_n \log_2 \Bigg (1 + 
\frac{ P^t_n/B_{\max}  }{ \gamma \sigma^2  || \mathcal{L} - \lim_{k\rightarrow \infty}  \mathcal{L}_n^k ||_2^{\alpha} + \beta \sigma^2 } \Bigg). \nonumber \\
\end{eqnarray}
Clearly $\widetilde{C}_S \in \mathcal{T}^{N_{BS}}_P $, $\mathcal{T}^{N_{BS}}_L$ is a Chebyshev set. Using Theorem~\ref{thm:optimal_para}, 
there exists an optimal BS locations configuration with the given bandwidth and transmit power.

Therefore, the JS divergence can be minimized by alternatively optimizing the BS location, the bandwidth, and the transmit power.

\section{Proof of Lemma~\ref{lem:bounded_problem}}
\label{appendix:bounded_problem}
According to Theorem~\ref{thm:arbitrarily_approx} and Theorem~\ref{thm:optimal_para},
there exists  $\{\Bar{\mathcal{L}_n}\}, \{\Bar{B_n}\}, \{\Bar{P^{t}_n}\} = \underset{\{\mathcal{L}_n\}, \{B_n\},\{P^t_n\}}{\arg\min} \xi\left(C_S, D_{T}\right)$. Note  $C_S(\{\Bar{\mathcal{L}_n}\}, \{\Bar{B_n}\}, \{\Bar{P^{t}_n}\})$ as $\Bar{C}_S$, the lower bound of optimal IREE is given by
\begin{align}
\eta^{\star}_{IREE} &\overset{(a)}{\geq} \underset{\{\mathcal{L}_n\}, \{B_n\},\{P^t_n\}}{\max} \frac{\min\{C_{Tot},D_{Tot}\}\left[1 - \xi\left(C_{T}, D_{T}\right)\right]}{\lambda P_{\max} + N_{BS} P^{c}} 
\nonumber \\
&= \frac{D_{Tot} \left[ 1 - \xi\left(\Bar{C}_S, D_{T}\right) \right]}{\lambda P_{\max} + N_{BS} P^{c}}, 
\end{align}
where step (a) holds according to the maximum transmit power constraint \eqref{constrain:max_power}.

On the other hand, with $C_{Tot} \geq D_{Tot}$ and minimized JS divergence $\xi\left(\Bar{C}_S, D_{T}\right)$, the upper bound of the IREE should be obtained when the minimum power consumption is met while satisfying the total traffic requirement $D_{Tot}$. We first consider a single BS scenario, where a single BS is located at the central location of area, $\mathcal{L}_{\mathcal{A}}$. By applying the Jensen's inequality, we have $C_{Tot} \geq  V_{\mathcal{A}} B_{\max} \log_2 \left (1 + \frac{ P^t/(\sigma^2 B_{\max}) }{ \frac{1}{V_{\mathcal{A}} }  \iint_{\mathcal{A}} L(\mathcal{L}, \mathcal{L}_{\mathcal{A}}) \textrm{d}\mathcal{L} } \right) \geq D_{Tot}$, where $V_{\mathcal{A}}$ is the volume of area $\mathcal{A}$. Hence, we have $P^{t} \geq \frac{B_{\max} \sigma^2 \iint_{\mathcal{A}} L(\mathcal{L}, \mathcal{L}_{\mathcal{A}}) \textrm{d}\mathcal{L}}{V_{\mathcal{A}}}   \left ( 2^{\frac{D_{Tot}}{V_{\mathcal{A}} B_{\max}}} - 1 \right ) \triangleq P^{t}_D.$

In the case where more than one BS is deployed, we can demonstrate that the power consumption required is greater than that of the single BS scenario mentioned earlier. To illustrate this, let us consider the best case where all the BSs are positioned at the central location $\mathcal{L}_{\mathcal{A}}$ to maximize the capacity. By applying the Jensen's inequality on concave function $C_S(P^t_n)$, we have $\iint_{\mathcal{A}} \sum_{n=1}^{N_{BS}} 
C_S(\mathcal{L}) \textrm{d}\mathcal{L} \leq \iint_{\mathcal{A}} B_{\max} \log_2 \left (1 + \frac{ \sum_{n=1}^{N_{BS}} \frac{B_n}{B_{\max}}  P^t_n}{ \sigma^2 B_{\max} L(\mathcal{L}, \mathcal{L}_{\mathcal{A}}) } \right) \textrm{d}\mathcal{L}$. Hence, the total capacity of this $N_{BS}$ BS is less than single BS with transmit power $\sum_{n=1}^{N_{BS}} \frac{B_n}{B_{\max}}  P^t_n$. 

Therefore, in order to achieve the same $D_{Tot}$, the power consumption needed in multi-BS scenario is larger than single BS scenario, $ \sum_{n=1}^{N_{BS}} P^t_n \geq P^t \geq P^{t}_D$. Hence we can obtain the upper bound of IREE as follows,
\begin{align} 
\label{eq:}
\eta^{\star}_{IREE} & \leq \underset{\{\mathcal{L}_n\}, \{B_n\},\{P^t_n\}}{\textrm{maximize}} \frac{\min\{C_{Tot},D_{Tot}\}\left[1 - \xi\left(\Bar{C}_S, D_{T}\right) \right]}{P_T} \nonumber \\
&\leq \frac{D_{Tot}\left[1 - \xi\left(\Bar{C}_S, D_{T}\right)\right]}{\lambda P^{t}_D + N_{BS} P^{c}}.
\end{align}

\section{Proof of Claim~\ref{cla:not_smooth}}
\label{appendix:not_smooth}

First of all, let us consider the smoothness of $C_S$ with respect to $P^t_n$. From the explicit expression of $C_S$ \eqref{eqn:def_Cs}, we have $\frac{\partial C_S }{ \partial P^t_n } = \frac{B_n}{\ln{2} ( P^t_n + \Lambda(\mathcal{L}_n, \mathcal{L}) ) }$ and $\frac{\partial^2 C_S }{ \partial (P^t_n)^2 } = -\frac{B_n}{\ln{2} ( P^t_n + \Lambda(\mathcal{L}_n, \mathcal{L}) )^2 }$, where $\Lambda(\mathcal{L}_n, \mathcal{L}) = B_{\max} \gamma \sigma^2  || \mathcal{L} - \mathcal{L}_n ||_2^{\alpha} + B_{\max} \beta \sigma^2$. According to \cite{zhang2019gradient}, $C_S$ is $(L_0,L_1)$ smoothness if and only if there exist $L^P_0$ and $L^P_1$ such that $\left \Vert \frac{\partial^2 C_S }{ \partial (P^t_n)^2 } \right \Vert \leq L^P_0 + L^P_1 \left \Vert \frac{\partial C_S }{ \partial P^t_n } \right \Vert$. Substitute the first order and the second order derivative into above inequality, we can obtain that $P^t_n \geq \frac{2 B_n}{ \sqrt{(B_n L^P_1)^2 + 4 \ln{2} L^P_0 } + B_n L^P_1 } - \Lambda(\mathcal{L}_n, \mathcal{L})$. Substitute into \eqref{constrain:max_power}, we have
\begin{eqnarray} 
\label{eq:cs_smooth_condition} 
\sum_{n=1}^{N_{BS}} \Lambda(\mathcal{L}_n, \mathcal{L}) &\geq& \sum_{n=1}^{N_{BS}}
\frac{2 B_n}{ \sqrt{(B_n L^P_1)^2 + 4 \ln{2} L^P_0 } + B_n L^P_1 } \nonumber \\
&-&  P_{\max},
\end{eqnarray}
which is the $(L_0,L_1)$ smoothness criteria of $C_S$ in one-shot training. This criteria tells us that every sample $\mathcal{L}$ must be far enough to all the BSs $\mathcal{L}_n$ to guarantee the $(L_0,L_1)$ smoothness, which is usually not met in practical scenarios. 

Based on the above analysis of $C_S$, we now discuss the smoothness of the loss function $L_{err}$.
To simplify the analysis, we consider the case that $L_{err}$ is a convex function of $C_S$. Therefore, for any $L^P_0 $ and $L^P_1$, we have

\begin{eqnarray} 
\label{eq:} 
\left \Vert \frac{\partial^2 L_{err} }{ \partial (P^t_n)^2 } \right \Vert &=& \left \Vert \frac{\partial L_{err} }{ \partial C_S } \cdot \frac{\partial^2 C_S }{ \partial (P^t_n)^2 } + \frac{\partial^2 L_{err} }{ \partial (C_S)^2 } \cdot \left( \frac{\partial C_S }{ \partial P^t_n } \right )^2 \right \Vert \nonumber \\
&\overset{(a)}{\geq}& \left \Vert \frac{\partial L_{err} }{ \partial C_S } \right \Vert  \cdot \left \Vert \frac{\partial^2 C_S }{ \partial (P^t_n)^2 } \right \Vert \nonumber \\
&\overset{(b)}{>}& \left \Vert \frac{\partial L_{err} }{ \partial C_S } \right \Vert  \cdot \left (L^P_0 + L^P_1 \left \Vert \frac{\partial C_S }{ \partial P^t_n } \right \Vert  \right  ) \nonumber \\
&=& L^P_1 \left \Vert \frac{\partial L_{err} }{ \partial P^t_n } \right \Vert  + L^P_0 \left \Vert \frac{\partial L_{err} }{ \partial C_S } \right \Vert,
\end{eqnarray}
where step (a) is obtained according the convexity of $L_{err}$ and step (b)  holds when $C_S$ does not satisfy the smoothness condition \eqref{eq:cs_smooth_condition}. Hence, $L_{err}$ does not satisfy the $(L_0,L_1)$ smoothness in one-shot training strategy, which leads to the VGP according to \cite{wang2022provable}.

In the proposed two stage learning, we do not need to satisfy the \eqref{constrain:max_power} in the first training stage. This ensures the smoothness of $C_S$, therefore prevents the vanishing gradient problem.

\section{Proof of Theorem~\ref{thm:conver_analy}}
\label{appendix:conver_analy}

According to \cite{wang2022provable}, the Adam optimizer converges to a bounded region if the loss function $L_{err}$ satisfies $(L_0,L_1)$ smoothness. Hence, after enough training epochs $N_{epoch}$, we have

\begin{eqnarray}
&& E^{(k)}(\{\mathcal{L}_n^{(k-1)}\}, \{B_n^{(k-1)}\},\{P^{t,(k-1)}_n\} ) \nonumber \\
&& + \bm{\omega} \Omega(\{\mathcal{L}_n^{(k-1)}\}, \{B_n^{(k-1)}\},\{P^{t,(k-1)}_n\} ) \nonumber \\
&\overset{(a)}{=}& \bm{\omega} \Omega(\{\mathcal{L}_n^{(k-1)}\}, \{B_n^{(k-1)}\},\{P^{t,(k-1)}_n\} ) \nonumber \\
&\overset{(b)}{\geq}& E^{(k)}(\{\mathcal{L}_n^{(k)}\}, \{B_n^{(k)}\},\{P^{t,(k)}_n\} ) \nonumber \\
&& + \bm{\omega} \Omega(\{\mathcal{L}_n^{(k)}\}, \{B_n^{(k)}\},\{P^{t,(k)}_n\} ) \nonumber \\
&=& - \min \{C_{Tot}(\{\mathcal{L}_n^{(k)}\}, \{B_n^{(k)}\},\{P^{t,(k)}_n\} ), D_{Tot}\} \nonumber \\
&& \times [1 - \xi(\{\mathcal{L}_n^{(k)}\}, \{B_n^{(k)}\},\{P^{t,(k)}_n\} )] \nonumber \\
&& + \eta^{(k)}_{IREE} P_T(\{\mathcal{L}_n^{(k)}\}, \{B_n^{(k)}\},\{P^{t,(k)}_n\} ) \nonumber \\
&& + \bm{\omega} \Omega(\{\mathcal{L}_n^{(k)}\}, \{B_n^{(k)}\},\{P^{t,(k)}_n\} ),
\label{eqn:}
\end{eqnarray}
where step (a) holds according to the definition of $\eta^{(k)}_{IREE}$, i.e. \eqref{eqn:iree_update}, and step (b) holds due to the convergence of RBF network. Therefore, we have

\begin{eqnarray}
&& \eta^{k}_{IREE}  \nonumber \\
&\leq& \min \{C_{Tot}(\{\mathcal{L}_n^{(k)}\}, \{B_n^{(k)}\},\{P^{t,(k)}_n\} ), D_{Tot}\}  \nonumber \\
&& \times [1 - \xi(\{\mathcal{L}_n^{(k)}\}, \{B_n^{(k)}\},\{P^{t,(k)}_n\} )]  \nonumber \\
&& \times \frac{1}{P_T(\{\mathcal{L}_n^{(k)}\}, \{B_n^{(k)}\},\{P^{t,(k)}_n\} )} \nonumber \\
&&+ \bm{\omega} [\Omega(\{\mathcal{L}_n^{(k-1)}\}, \{B_n^{(k-1)}\},\{P^{t,(k-1)}_n\} ) \nonumber \\
&&- \Omega(\{\mathcal{L}_n^{(k)}\}, \{B_n^{(k)}\},\{P^{t,(k)}_n\} )] \nonumber \\
&& \times \frac{1 }{P_T(\{\mathcal{L}_n^{(k)}\}, \{B_n^{(k)}\},\{P^{t,(k)}_n\} )} \nonumber \\
&\triangleq& \eta^{(k+1)}_{IREE} +\Delta \Omega^k,
\label{eqn:}
\end{eqnarray}
where $\Delta \Omega^k = \bm{\omega} [\Omega(\{\mathcal{L}_n^{(k-1)}\}, \{B_n^{(k-1)}\},\{P^{t,(k-1)}_n\} ) - \Omega(\{\mathcal{L}_n^{(k)}\}, \{B_n^{(k)}\},\{P^{t,(k)}_n\} )]/P_T(\{\mathcal{L}_n^{(k)}\}, \{B_n^{(k)}\},\{P^{t,(k)}_n\} )$. 

Since $ \lim_{k \rightarrow \infty} L_{err}^{(k)}(\{\mathcal{L}_n\}, \{B_n\},\{P^t_n\};\bm{\omega}) = 0$, for any $\epsilon > 0$, there exists a $K \geq 0$, such that for any $k \geq K$, 
$\Delta \Omega^k < \epsilon$. With $\epsilon \ll \eta^{(k)}_{IREE}$, we have $\eta^{(k+1)}_{IREE} \geq \eta^{(k)}_{IREE}$.

Also, we shall prove that $ \lim_{k \rightarrow \infty} \eta^{(k)}_{IREE} = \eta^{\star}_{IREE}$. Suppose $ \lim_{k \rightarrow \infty} \eta^{(k)}_{IREE} = \hat{\eta}_{IREE} \neq \eta^{\star}_{IREE}$, we must have $\hat{\eta}_{IREE} < \eta^{\star}_{IREE}$. Note  function $\mathcal{F}(\eta_{IREE})$ as $\mathcal{F}(\eta_{IREE}) = \underset{\{\mathcal{L}_n\}, \{B_n\},\{P^t_n\}}{\textrm{maximize}}  \Big\{ \min\{C_{Tot},D_{Tot}\}\big[1 - \xi\left(C_{T}, D_{T}\right)\big] - \eta_{IREE} P_T \Big\}$, then $\mathcal{F}(\eta_{IREE})$ is a monotonic decreasing function as given in \cite{dinkelbach1967nonlinear}. If Algorithm~\ref{alg:Dinkelbach} is converged, then according to the terminal condition we have $\mathcal{F}(\hat{\eta}_{IREE}) = 0$. On the other hand, we have  $\mathcal{F}(\eta^{\star}_{IREE}) = 0$ according to Lemma~\ref{lem:optimal_condition}. Thus, this contradicts with the fact that $\hat{\eta}_{IREE} < \eta^{\star}_{IREE}$. Hence, it follows that  $ \lim_{k \rightarrow \infty} \mathcal{F}(\eta^{(k)}_{IREE}) = \mathcal{F}(\eta^{\star}_{IREE})$ and  $ \lim_{k \rightarrow \infty} \eta^{(k)}_{IREE} = \eta^{\star}_{IREE}$.

\section{Proof of Lemma~\ref{lem:iree_gap}}
\label{appendix:iree_gap}
According to the triangle inequality of the JS divergence \cite{osan2018monoparametric}, the gap between $\xi^{T} $ and $\xi^{S} $ is given by $|\xi^{T} - \xi^{S}| = |\xi\left(C_{S}, D_{T}\right) - \xi\left(C_{T}, D_{T}\right)| \leq \xi\left(C_{S}, C_{T}\right) \triangleq \xi^{S,T}$. If $ C^{S}_{Tot} > D_{Tot}$, the optimality gap $\Delta \eta = \frac{ D_{Tot} |\xi\left(C_{S}, D_{T}\right) - \xi\left(C_{T}, D_{T}\right)|}{P_T} \leq \frac{ D_{Tot} \xi^{S,T}}{P_T} $. If $ C^{S}_{Tot} \leq D_{Tot}$, the optimality gap $\Delta \eta$ is given by,
\begin{eqnarray}
\label{eq:}
\Delta \eta &\overset{(a)}{\leq}& \frac{ |C^{T}_{Tot}- C^{S}_{Tot} \frac{1 - \xi^{S}}{1 - \xi^{T}} |}{P_T}  \nonumber \\
&\overset{(b)}{\leq} & \frac{ (1 - \xi^{S})(C^{T}_{Tot}- C^{S}_{Tot} ) +  \xi^{S,T} C^{T}_{Tot}   }{(1 - \xi^{S} +  \xi^{S,T} ) P_T}. 
\label{eq:} 
\end{eqnarray}
In the above equations, step (a) is obtained since $1 - \xi^{T} \leq 1$, and step (b) is obtained according to the triangle inequality.
\end{appendices}

\bibliographystyle{IEEEtran}
\bibliography{IEEEfull,references}

\begin{IEEEbiography}[{\includegraphics[width=1in,height=1.25in,clip,keepaspectratio]{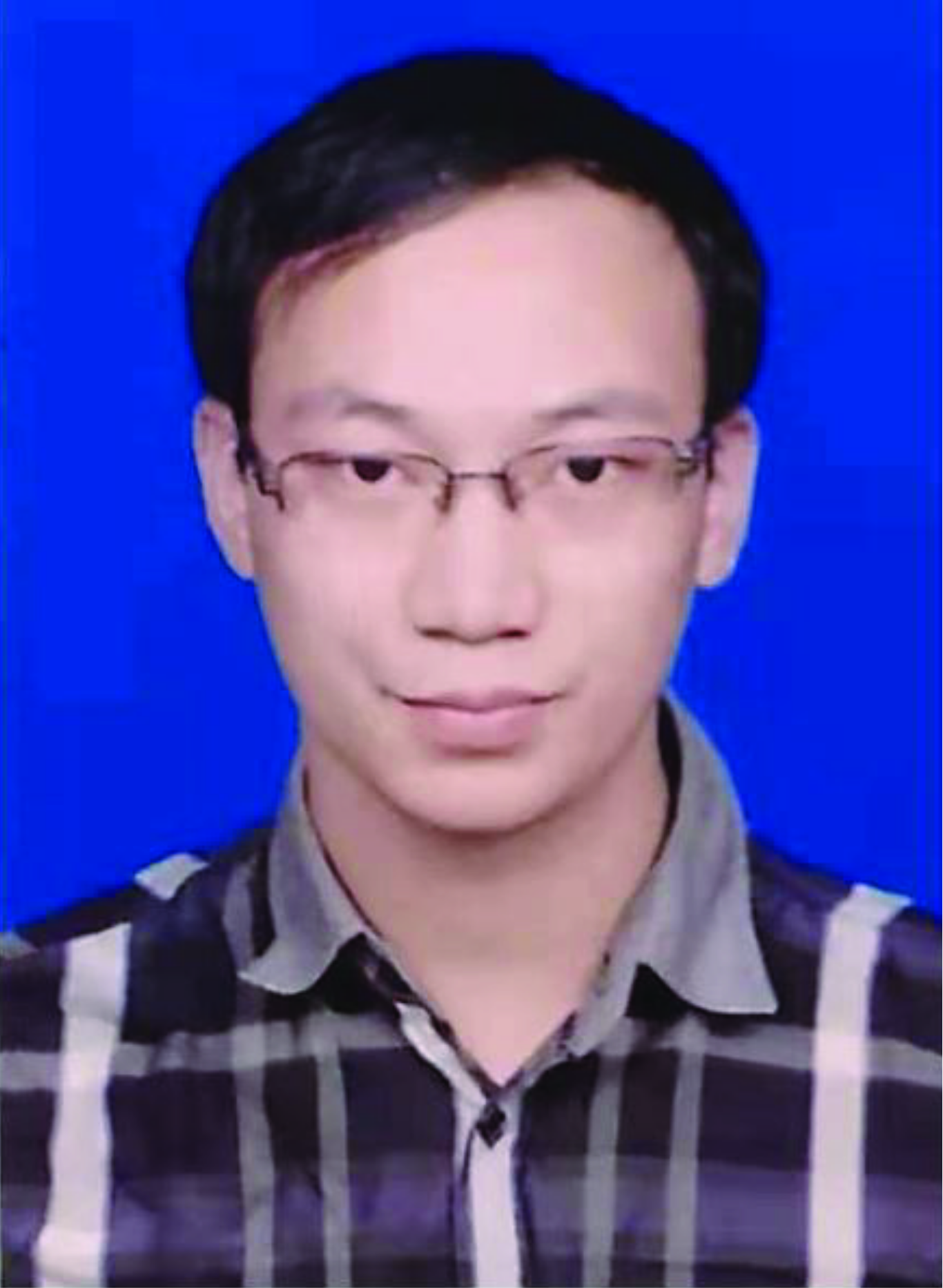}}]{Tao Yu}
received the B.E. and M.E. degrees from the School of Communication and Information Engineering, Shanghai University, in 2018 and 2021, respectively, where he is currently pursuing the Ph.D. degree with the School of Communication and Information Engineering. His research fields include energy-efficient communication networks, machine learning, and deep learning in wireless network.
\end{IEEEbiography}

\begin{IEEEbiography}[{\includegraphics[width=1in,height=1.25in,clip,keepaspectratio]{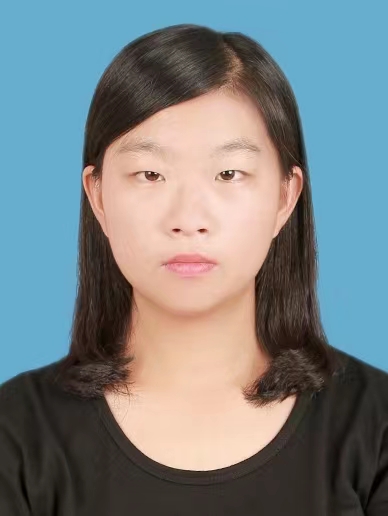}}]{Pengbo Huang}
received the B.E. degree from the School of Communication and Information Engineering, Shanghai University in 2022, where she is currently pursuing the M.E. degree with the School of Communication and Information Engineering. Her research fields include green communication networks, machine learning and deep learning.
\end{IEEEbiography}

\begin{IEEEbiography}[{\includegraphics[width=1in,height=1.25in,clip,keepaspectratio]{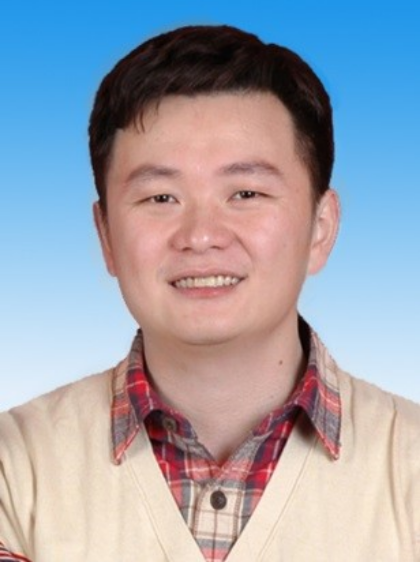}}]{Shunqing Zhang}
(Senior Member, IEEE) received the B.S. degree from the Department of Microelectronics, Fudan University, Shanghai, China, in 2005, and the Ph.D. degree from the Department of Electrical and Computer Engineering, Hong Kong University of Science and Technology, Hong Kong, in 2009. He was with the Communication Technologies Laboratory, Huawei Technologies, as a Research Engineer and then a Senior Research Engineer from 2009 to 2014, and a Senior Research Scientist of Intel Collaborative Research Institute on Mobile Networking and Computing, Intel Labs from 2015 to 2017. Since 2017, he has been with the School of Communication and Information Engineering, Shanghai University, Shanghai, China, as a Full Professor. His current research interests include energy efficient 5G/5G+ communication networks, hybrid computing platform, and joint radio frequency and baseband design. He has published over 60 peer-reviewed journal and conference papers, as well as over 50 granted patents. He has received the National Young 1000-Talents Program and won the paper award for Advances in Communications from IEEE Communications Society in 2017.
\end{IEEEbiography}

\begin{IEEEbiography}[{\includegraphics[width=1in,height=1.25in,clip,keepaspectratio]{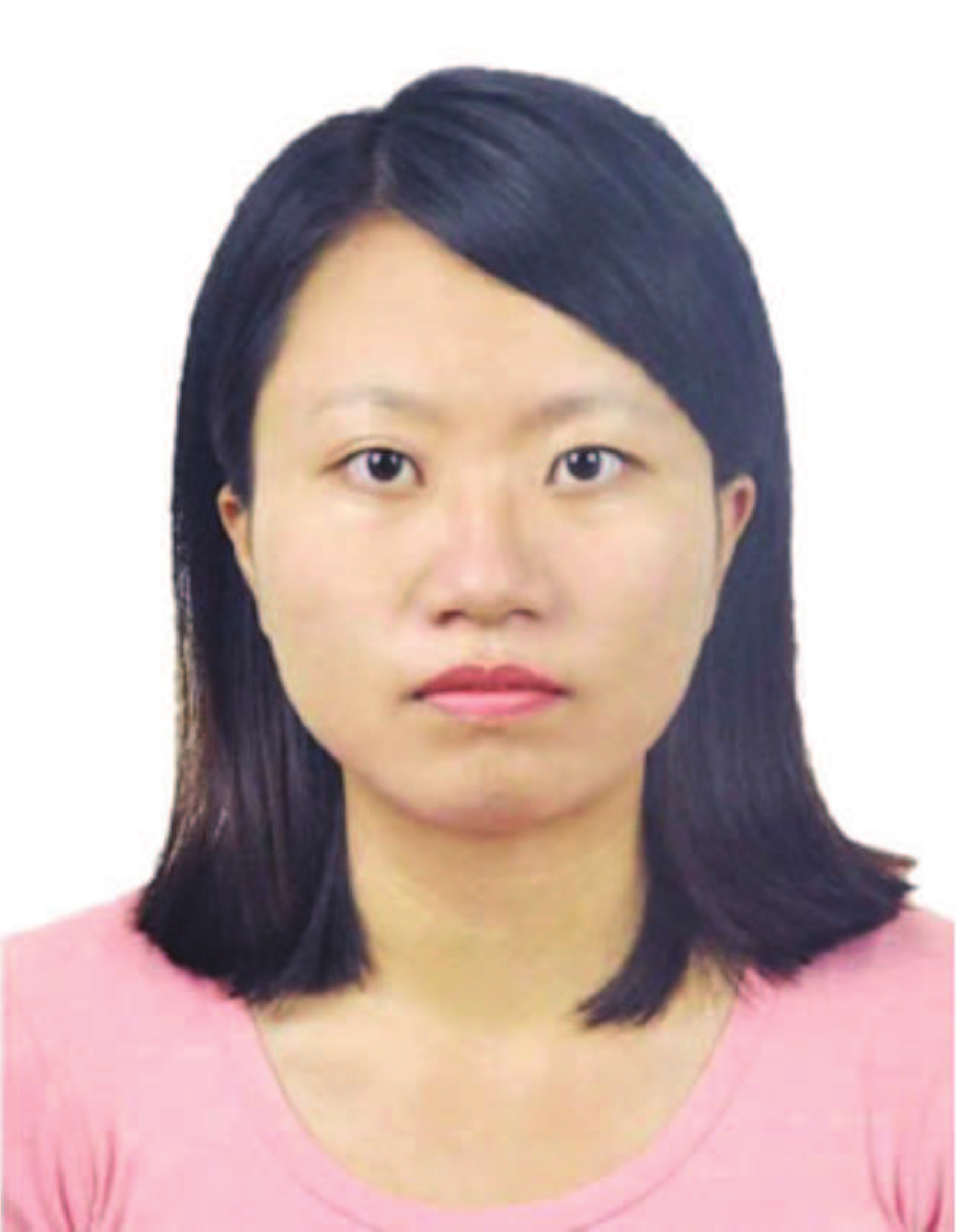}}]{Xiaojing Chen}
(Member, IEEE) received the B.E. degree in communication science and engineering and the Ph.D. degree in electromagnetic field and microwave technology from Fudan University, China, in 2013 and 2018, respectively, and the Ph.D. degree in engineering from Macquarie University, Australia, in 2019. She is currently an associate professor with Shanghai University, China. Her research interests include wireless communications, energy-efficient communications, stochastic network optimization, and network functions virtualization.
\end{IEEEbiography}

\begin{IEEEbiography}[{\includegraphics[width=1in,height=1.25in,clip,keepaspectratio]{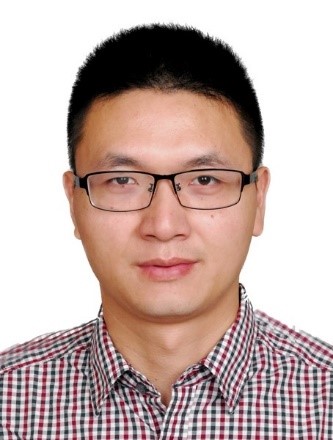}}]{Yanzan Sun}
(Member, IEEE) received the Ph.D. degree in Control Theory and Control Engineering from Tongji University, China, in 2011. In 2009, he was awarded a National government study abroad scholarship to pursue his research as a visiting student in Electrical Engineering Department at Columbia University. From 2011 to 2012, he was a Research Engineer at Huawei Technologies Co., Ltd. where he researched enhanced Inter-cell Interference Coordination (eICIC) in 3GPP Long Term Evolution (LTE) systems. He is currently an associate professor in the School of Communications and information engineering at Shanghai University. His research interests include key technologies of 5G, wireless resource management, and green communications. 
\end{IEEEbiography}

\begin{IEEEbiography}[{\includegraphics[width=1in,height=1.25in,clip,keepaspectratio]{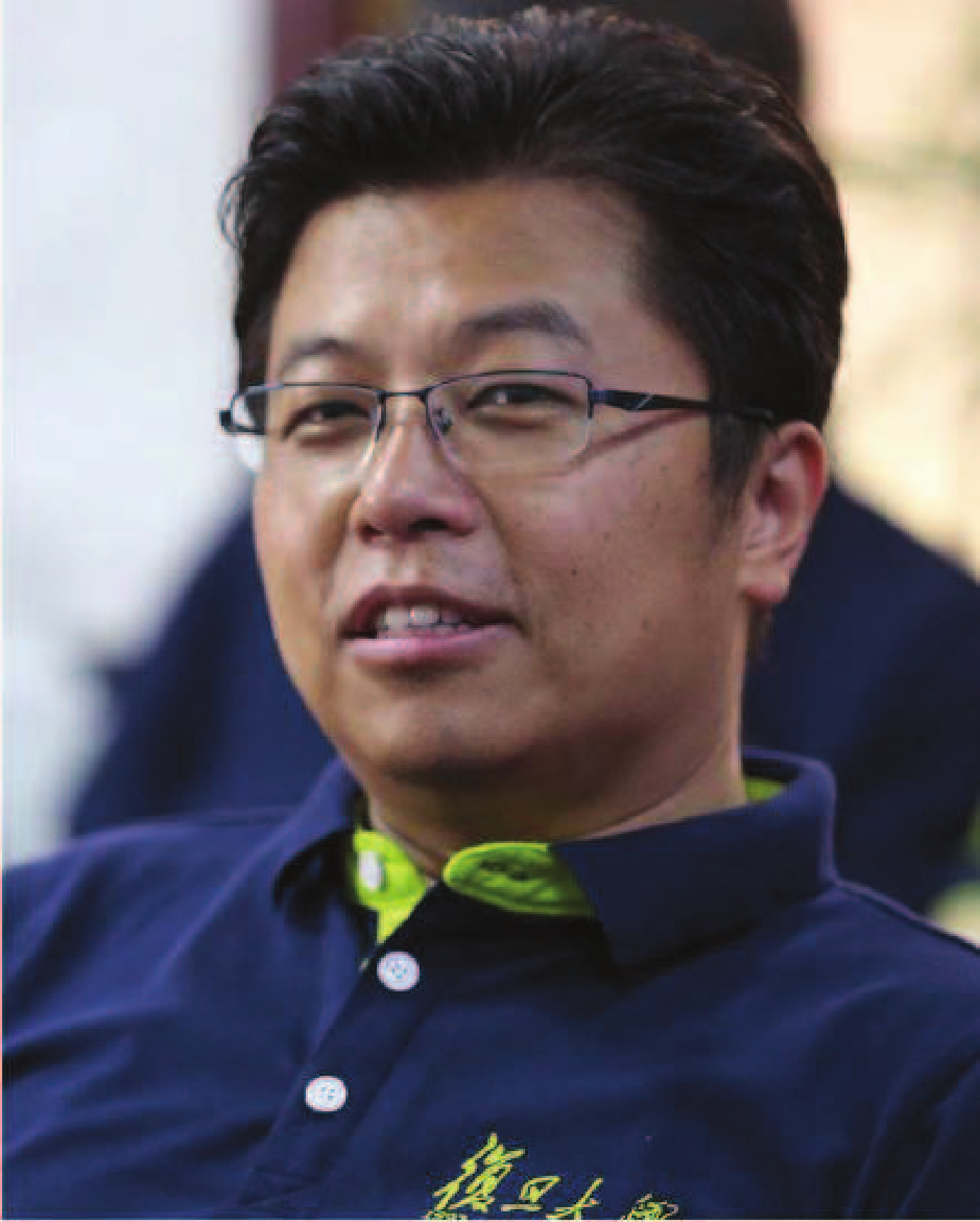}}]{Xin Wang}
(Fellow, IEEE) received the B.Sc. and M.Sc. degrees in electrical engineering from Fudan University, Shanghai, China, in 1997 and 2000, respectively, and the Ph.D. degree in electrical engineering from Auburn University, Auburn, AL, USA, in 2004. From September 2004 to August 2006, he was a Post-Doctoral Research Associate with the Department of Electrical and Computer Engineering, University of Minnesota, Minneapolis. In August 2006, he joined the Department of Electrical Engineering, Florida Atlantic University, Boca Raton, FL, USA, as an Assistant Professor, then was promoted to a Tenured Associate Professor in 2010. He is currently a Distinguished Professor and the Chair of the Department of Communication Science and Engineering, Fudan University. His research interests include stochastic network optimization, energy-efficient communications, cross-layer design, and signal processing for communications. He is an IEEE Distinguished Lecturer of the Vehicular Technology Society. He served as a Senior Area Editor for IEEE Transactions on Signal Processing, an Editor for IEEE Transactions on Vehicular Technology, and an Editor for IEEE Transactions on Wireless Communications
\end{IEEEbiography}

\end{document}